\documentclass[twocolumn,onecolappendix]{aastex62}
\usepackage{amsmath}

\shortauthors{Sheehan et al.}
\shorttitle{Dynamical Masses with ALMA and Gaia}

\begin{document}

\title{\bf High Precision Dynamical Masses of Pre-Main Sequence Stars with ALMA and Gaia}

\author{Patrick D. Sheehan}
\affiliation{National Radio Astronomy Observatory, 520 Edgemont Rd., Charlottesville, VA 22901, USA}
\affiliation{Homer L. Dodge Department of Physics \& Astronomy, University of Oklahoma, 440 W. Brooks St., Norman, OK 73021, USA}

\author{Ya-Lin Wu}
\altaffiliation{51 Pegasi b Fellow}
\affiliation{McDonald Observatory and the Department of Astronomy, University of Texas at Austin, Austin, TX 78712, USA}

\author{Josh A. Eisner}
\affiliation{University of Arizona Department of Astronomy and Steward Observatory, 933 North Cherry Ave., Tucson, AZ 85721, USA}

\author{John Tobin}
\affiliation{National Radio Astronomy Observatory, 520 Edgemont Rd., Charlottesville, VA 22901, USA}
\affiliation{Homer L. Dodge Department of Physics \& Astronomy, University of Oklahoma, 440 W. Brooks St., Norman, OK 73021, USA}

\begin{abstract}
The Keplerian rotation in protoplanetary disks can be used to robustly measure stellar masses at very high precision if the source distance is known. We present Atacama Large Millimeter/submillimeter Array (ALMA) observations of spatially and spectrally resolved $^{12}$CO (2--1) emission towards the disks around 2MASS J16262774--2527247 (the tertiary companion to ROXs 12 at 5100 au), CT Cha, and DH Tau. We employ detailed modeling of the Keplerian rotation profile, coupled with accurate distances from $Gaia$, to directly measure the stellar masses with $\sim$2\% precision. We also compare these direct mass measurements with the masses inferred from evolutionary models, determined in a statistically rigorous way. We find that 2MASS J16262774--2527247 has a mass of $0.535^{+0.006}_{-0.007}~M_\sun$ and CT Cha has a mass of $0.796^{+0.015}_{-0.014}~M_\sun$, broadly consistent with evolutionary models, although potentially significant differences remain. DH Tau has a mass of $0.101^{+0.004}_{-0.003}~M_\sun$, but it suffers from strong foreground absorption that may affect our mass estimate. The combination of ALMA, $Gaia$, and codes like \texttt{pdspy}, presented here, can be used to infer the dynamical masses for large samples of young stars and substellar objects, and place constraints on evolutionary models.
\end{abstract}

\keywords{protoplanetary disks --- stars: fundamental parameters --- stars: pre-main sequence --- radio lines: stars --- parallaxes}

\section{Introduction}

Protoplanetary disks are found nearly ubiquitously around pre-main sequence stars \citep[e.g.,][]{Hernandez2008}, and studies of these disks are crucial to understanding the environment in which stars and planets form. Now, thanks to the revolutionary power of the Atacama Large Millimeter/submillimeter Array (ALMA), these disks have been studied in great detail. One notable development has been the unprecedented sensitivity to molecular line emission, which was inaccessible by previous generations of millimeter interferometers for all but the brightest sources\citep[e.g.,][]{Simon2000}. Access to gas emission has opened up a number of new avenues of study of protoplanetary disks, including their chemistry \citep[e.g.,][]{Oberg2015,Schwarz2016}, direct measurements of gas disk sizes and masses \citep[e.g.,][]{Williams2014,Ansdell2016,Ansdell2018,Miotello2017}, as well as constraints on the turbulence in disks \citep{Flaherty2015,Flaherty2017,Flaherty2018}.

While there has been a significant effort toward studies of disk structure, the detection of gas emission has also created an opportunity to study the central star. Detailed radiative transfer modeling of the Keplerian motion in spatially and spectrally resolved disks has long been used as a tool for directly measuring stellar masses (e.g., \citealt{Dutrey1994,Simon2000,Dutrey2003,Czekala2015,Czekala2016,MacGregor2017,Ricci2017,Czekala2017,Wu2017b,Simon2017}). However, the precision of these measurements is limited almost entirely by the uncertainty in the distance to the star, typically $>$10\%, thereby limiting their usefulness for constraining evolutionary models. Although VLBI has produced precise distance measurements for some sources, potentially enabling precise mass measurements \citep[e.g.,][]{Simon2017}, care must be taken when extrapolating those distances (and distance uncertainties) to the star-forming region as a whole \citep[e.g.,][]{OrtizLeon2017}. In the era of precision distance measurements with $Gaia$, however, dynamical masses should be measurable with significantly improved precision for large samples of pre-main sequence stars.

\begin{deluxetable*}{lcccccc}
\tablenum{1}
\tablecaption{Log of ALMA Observations}
\label{table:obs}
\tablehead{\colhead{Source} & \colhead{Observation Date} & \colhead{Baselines} & \colhead{Integration Time} & \colhead{Beam} & \colhead{RMS} & \colhead{Calibrators} \\
\colhead{} & \colhead{(UT)} & \colhead{(m)} & \colhead{(min)} & \colhead{} & \colhead{(mJy)} & \colhead{(Flux/Bandpass, Gain)}}
\startdata
DH Tau & Sep. 14 2016 & 15$\,$-$\,$3247 & 13 & 0.30" $\times$ 0.17" & 4 & J0510+1800, J0433+2905 \\
2M1626-2527 & Sep. 16 2016 & 15$\,$-$\,$3143 & 12 & 0.21" $\times$ 0.18" & 3.1 & J1517-2422, J1634-2058 \\
CT Cha & Sep. 27 2016 & 15$\,$-$\,$3247 & 13.5 & 0.27" $\times$ 0.14" & 3.8 & J1107-4449, J1058-8003 \\
\enddata
\end{deluxetable*}

This development comes at a welcome time as, while theoretical evolutionary tracks have progressed significantly over the last few decades, major discrepancies remain between the models and the direct measurements of stellar radii and masses \citep{Hillenbrand2004,Gennaro2012,Stassun2014,Bell2016}. Up to this point, constraints on pre-main sequence tracks have primarily come from eclipsing binaries, however these systems are relatively rare, with only $\sim$20 having accurate mass measurements \citep[e.g.,][]{Stassun2014}. Astrometry of young binary systems provides another avenue for measuring masses \citep[e.g.,][]{Rizzuto2016,Dupuy2016,Rodet2018}, but requires precise measurements over a number of epochs. These measurements are also limited, by definition, to binary stars.  On the contrary many (or most) studies of young stars focus on single stars, which are easier to study.  While it is typically assumed that relations between mass and temperature, radius, or other stellar parameters are common for binary and single stars, we must actually measure masses for a sample of single stars to test this assumption.

Constraints on evolutionary tracks are important for understanding fundamental stellar parameters \citep[e.g.,][]{Stassun2014,Bell2016}, but also because such models are often used to draw conclusions about the star and planet formation processes. Cluster ages inferred from evolutionary tracks have been used to place constraints on protoplanetary disk lifetimes \citep[e.g.,][]{Haisch2001,Hernandez2008}. More recently, a large amount of effort has gone into studying the disk-mass to stellar mass relation, how it evolves with time, and what that evolution means for disk evolution and planet formation \citep[e.g.,][]{Andrews2013,Pascucci2016,Barenfeld2016,Ansdell2016,Eisner2018}. However, if stellar properties are derived from pre-main sequence evolutionary tracks, they are always model-dependent \citep[e.g.,][]{Andrews2013,WardDuong2018}.

Here we demonstrate how ALMA $^{12}$CO(2--1) observations and precise distances with $Gaia$ can be used measure high precision ($\sim$2\%) dynamical masses for young stars, using observations of the three pre-main sequence stars 2MASS J16262774--2527247, CT Cha, and DH Tau. These sources were originally observed as part of a program to look for disks around young, accreting planetary-mass companions in both continuum and spectral line emission. The results of this program as it pertains to the companions are presented in \cite{Wu2017a,Wu2017b,Wu2017c}. The star 2MASS J16262774--2527247 (hereafter 2M1626--2527) was observed due to a mix-up propagated through the literature that identified its coordinates as those of the $\sim$17 $M_{\rm Jup}$ planetary-mass companion ROXs 12 B \citep{Bowler2017}. Interestingly, 2M1626--2527 is also shown to be a companion to ROXs 12 A \citep{Bowler2017}. Although these sources are not truly ``single", this technique can be equally well applied to any star, single or multiple, with a disk.

The paper is structured as follows. In Section \ref{section:observations} we describe the ALMA observations of our three targets. In Section \ref{section:modeling} we present our modeling analysis of the spectral line observations, including their mass measurements, using our new code \texttt{pdspy}\footnote{The \texttt{pdspy} code can be found here: https://github.com/psheehan/pdspy} \citep{Sheehan2018b}. Then, in Section \ref{section:evolutionary_track_masses} we use theoretical evolutionary tracks to calculate stellar masses for our sources, and compare with our dynamically measured masses in Sections \ref{section:results} and \ref{section:discussion}. Finally we provide extensive appendices to further document the modeling technique as well as the \texttt{pdspy} code. \texttt{pdspy} can handle both spectral line and continuum observations, and is now publicly available. The continuum portion of the code was used in \citet{Sheehan2017b} but the code was not formally introduced at that time.

\begin{figure*}[t]
\centering
\figurenum{1}
\includegraphics[width=7in]{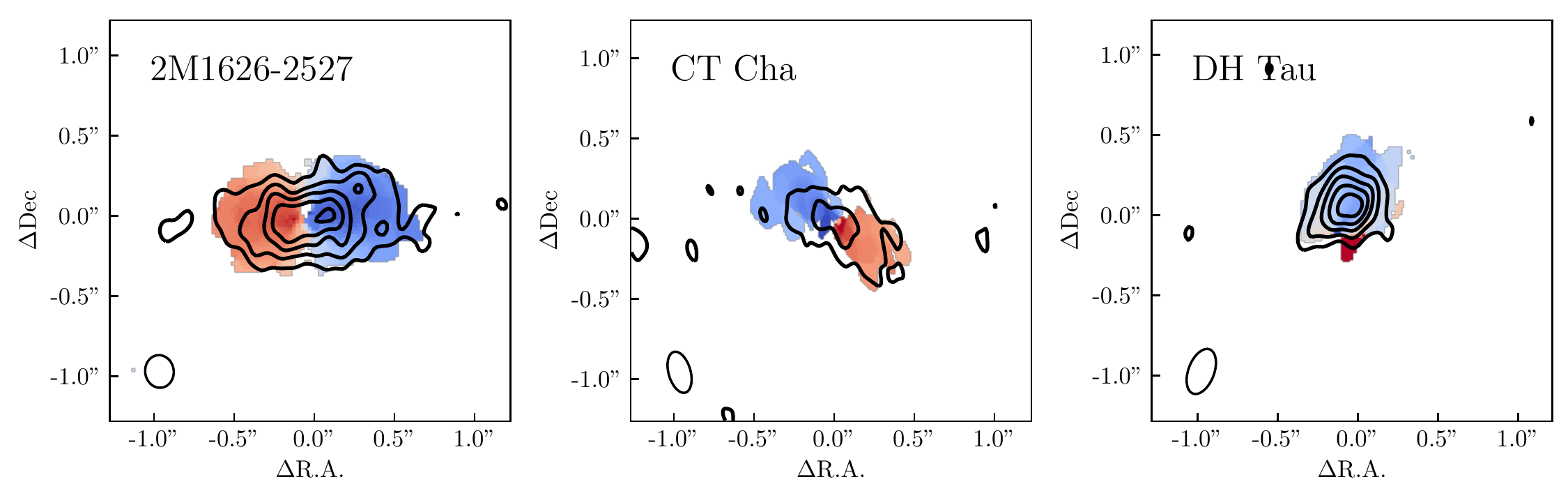}
\caption{CO (2--1) moment maps for all three of our sources. We show the moment 0 (integrated intensity) map as black contours and the moment 1 velocity map as a color image. Contours of the moment 0 map start at 2$\sigma$ and subsequent contours are in intervals of $3\sigma$. DH Tau shows a notable lack of redshifted emission, likely due to extinction by foreground clouds.}
\label{fig:moment_maps}
\end{figure*}

\section{Observations \& Data Reduction}
\label{section:observations}

Our targets were observed with ALMA Band 6 during Cycle 3, in September 2016. We include details of each individual observation in Table \ref{table:obs}. For each observation, the Band 6 receiver was set up with three basebands configured for wideband continuum observations, each with 2 GHz of bandwidth and centered at 233 GHz, 246 GHz, and 248 GHz. The final baseband was configured for $^{12}$CO (2--1) observations, with 3840 0.122 MHz channels centered at 230.538 GHz (0.32 km s$^{-1}$ velocity resolution; Hanning smoothed) in order to spectrally resolve the molecular line emission. In addition to observing our science targets, each track included time spent on quasars for the purposes of calibrating our data.
 
The data were calibrated using the ALMA Pipeline in the \texttt{CASA} software package \citep{McMullin2007}. Following the bandpass, flux, and gain calibrations, we employed a single iteration of phase-only self-calibration, with one solution per scan, to the continuum basebands and applied these phase solutions to our spectral line window. After calibration, the data were imaged and deconvolved using the spectral cube mode of the CASA {\it clean} task. The imaging was done using natural weighting to increase the sensitivity of the channel maps, as the emission is often faint, and with 0.4 km s$^{-1}$ channels. We provide additional details of the images in Table \ref{table:obs}, and for brevity we show moment 0 and 1 maps in Figure \ref{fig:moment_maps} and channel maps with a limited selection of channels in Figures \ref{fig:short_model_ROXs12}, \ref{fig:short_model_CTCha}, and \ref{fig:short_model_DHTau}. We do, however, show the full channel maps in Appendix \ref{section:channel_maps}. 

We note that there is a distinct lack of redshifted emission towards DH Tau from $\sim$6--7.2 km s$^{-1}$ (see Figure \ref{fig:moment_maps}, \ref{fig:short_model_DHTau}, and Appendix \ref{section:channel_maps}), with emission reappearing at 7.4 km s$^{-1}$. We suspect this lack of emission arises from extinction by molecular cloud in that velocity range, as in the case of GO Tau \citep{Schaefer2009}. We therefore exclude those channels from the channel map modeling described in the next section. The channels that were not considered are marked with an ``X" in any relevant figures. A similar effect (and resolution) were adopted by \citet{Czekala2016}.

\section{Analysis}
\label{section:analysis}

\subsection{Dynamical Stellar Masses with the \texttt{pdspy} Code}
\label{section:modeling}

%\subsection{}
%\label{section:modeling}

Keplerian rotation imprints a unique pattern in the disk kinematics that can be seen in channel maps (the so-called ``butterfly pattern''), and that pattern is strongly dependent on stellar mass. We generate synthetic Keplerian disk channel maps using radiative transfer modeling and fit these synthetic observations to the actual data using a Markov Chain Monte Carlo (MCMC) fitting routine. We describe our disk model, and our fitting procedure, in more detail below.

\begin{figure*}[t]
\centering
\figurenum{2}
\includegraphics[width=7in]{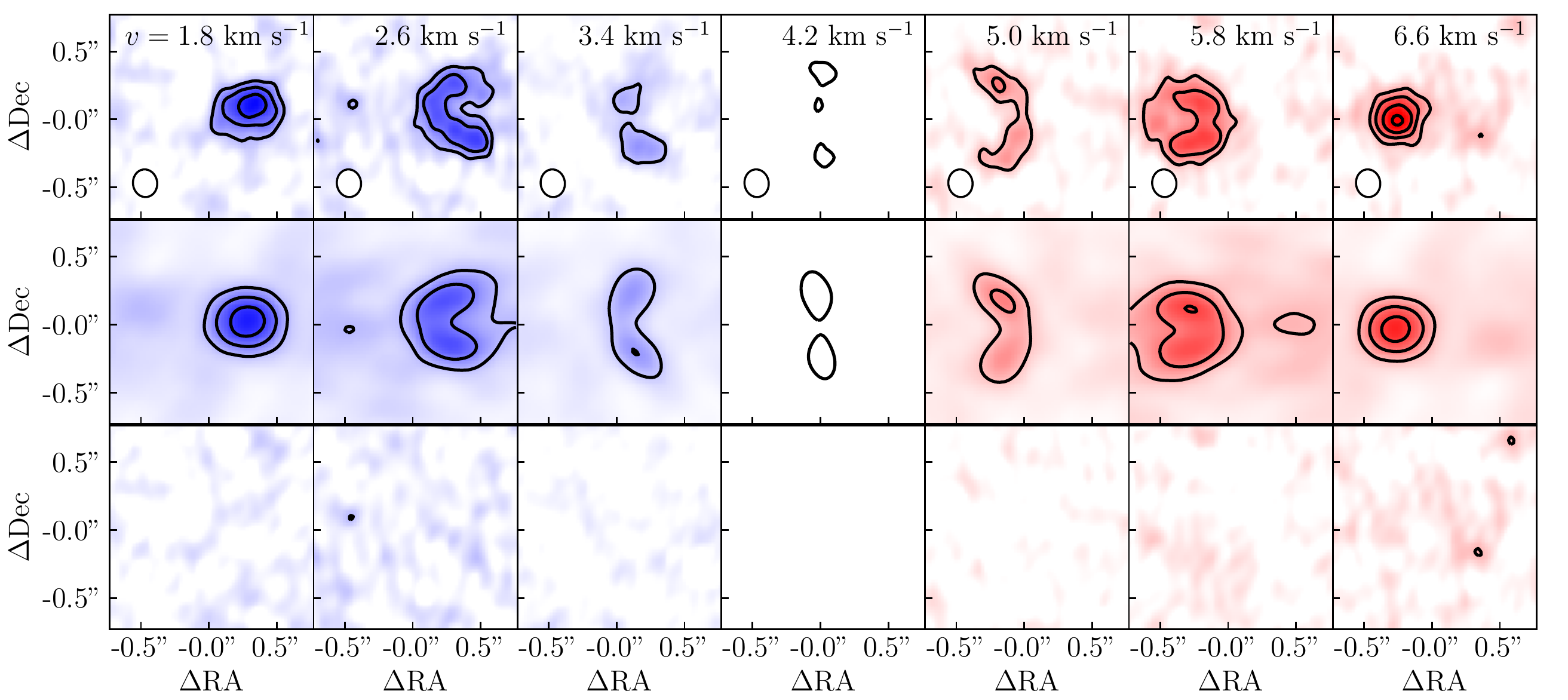}
\caption{Modeling results for 2M1626--2527. The first row shows our ALMA CO (2--1) channel maps, with contours starting at 3.5$\sigma$ and continuing in increments of 3$\sigma$. The second row shows the best fit model channel map images, and the third row shows the residuals, calculated in the Fourier plane and then Fourier transformed to produce an image. Here, for brevity, we show only a selection of channels, but the full channel maps can be found in Appendix \ref{section:channel_maps}}
\label{fig:short_model_ROXs12}
\end{figure*}

\begin{figure*}[t]
\centering
\figurenum{3}
\includegraphics[width=7in]{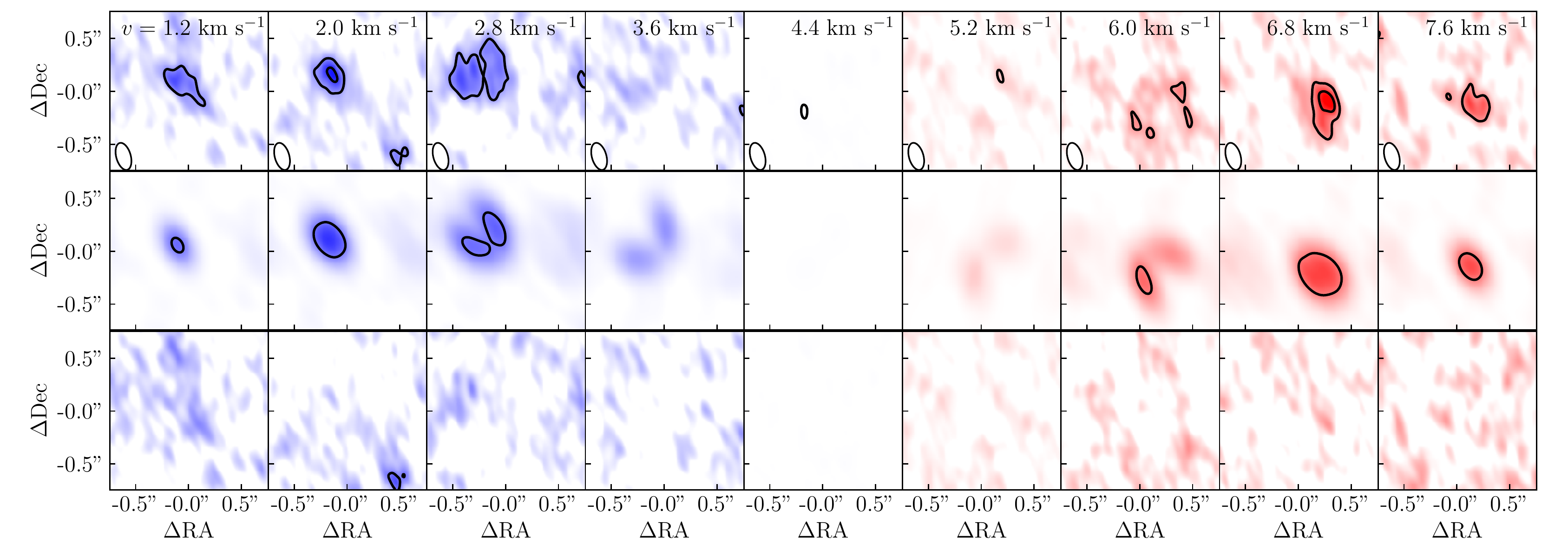}
\caption{The best fit model for CT Cha, including data, model, and residuals, shown in the same style as in Figure \ref{fig:short_model_ROXs12}.}
\label{fig:short_model_CTCha}
\end{figure*}

We assume a radial structure for our disk model that is motivated by models of viscous disk evolution \citep{LyndenBell1974}, with the surface density given by
\begin{equation}
\Sigma(r) = \, \Sigma_0 \left(\frac{r}{r_c}\right)^{-\gamma} \, \exp\left[-\left(\frac{r}{r_c}\right)^{-(2-\gamma)}\right],
\end{equation}
where $r$ is the stellocentric radius in cylindrical coordinates and $\gamma$ is the surface density power law exponent. $r_c$ is the radius beyond which the disk surface density is exponentially tapered, and serves as a proxy for the disk radius ($R_{disk} = r_c$) as the surface density profile has no hard outer limit. We assume that the disk is truncated at an inner radius of $R_{in}$. $\Sigma_0$ is a proportionality constant related to the total disk gas mass,
\begin{equation}
\Sigma_0 = \frac{(2-\gamma) M_d}{2 \pi r_c^2}.
\end{equation}
As total disk gas mass is easier to interpret, we leave it as a free parameter in place of $\Sigma_0$.

\begin{figure*}[t]
\centering
\figurenum{4}
\includegraphics[width=7in]{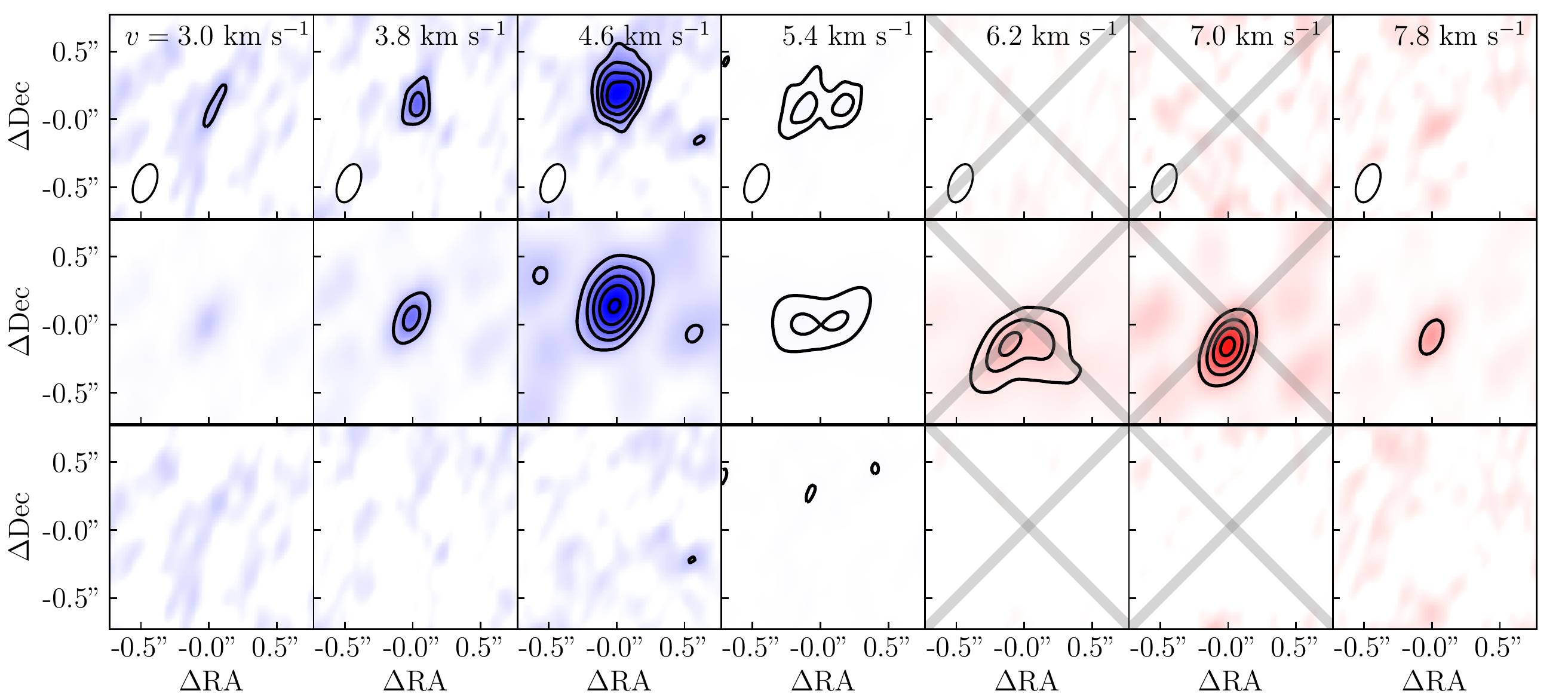}
\caption{The best fit model for DH Tau, including data, model, and residuals, shown in the same style as in Figure \ref{fig:short_model_ROXs12}. The channels with marked with an ``X" show a lack of redshifted emission, likely due to absorption by a foreground cloud. The ``X" indicates that these channels were excluded from our modeling.}
\label{fig:short_model_DHTau}
\end{figure*}

\begin{deluxetable*}{lcccccccccccccccc}
\tablecaption{Best-fit Model Parameters}
\tablenum{2}
\tabletypesize{\small}
\label{table:best_fits}
\tablehead{\colhead{Source} & \colhead{$M_{*}$} & \colhead{$\log M_{disk}$} & \colhead{$R_{in}$} & \colhead{$R_{disk}$} & \colhead{$\gamma$} & \colhead{$T_0$} & \colhead{$q$} & \colhead{$a_{turb}$} & \colhead{$v_{sys}$} & \colhead{$i$} & \colhead{$P.A.$} & \colhead{$d_{pc}$}\\ \colhead{ } & \colhead{$[M_{\odot}]$} & \colhead{$[M_{\odot}]$} & \colhead{[AU]} & \colhead{[AU]} & \colhead{ } & \colhead{[K]} & \colhead{} & \colhead{[km s$^{-1}$]} & \colhead{[km s$^{-1}$]} & \colhead{$[\,^{\circ}\,]$} & \colhead{$[\,^{\circ}\,]$} & \colhead{[pc]}}
\startdata
CT Cha & $0.796^{+0.015}_{-0.014}$ & $>-2.89$ & $<  6.1$ & $  9.3^{+ 12.0}_{-  5.1}$ & $ 0.9^{+ 0.2}_{- 0.4}$ & $  106^{+   22}_{-   14}$ & $0.44^{+0.04}_{-0.03}$ & $< 0.060$ & $4.608^{+0.014}_{-0.014}$ & $54.0^{+ 1.0}_{- 0.9}$ & $ 49.5^{+  0.6}_{-  0.6}$ & $191.8^{+  0.7}_{-  0.8}$ \\[2pt]
DH Tau & $0.101^{+0.004}_{-0.003}$ & $-4.54^{+0.32}_{-0.36}$ & $<  1.4$ & $< 17.1$ & $ 1.4^{+ 0.1}_{- 0.2}$ & $  237^{+   33}_{-   24}$ & $0.53^{+0.04}_{-0.04}$ & $< 0.187$ & $5.652^{+0.009}_{-0.010}$ & $48.4^{+ 1.4}_{- 1.5}$ & $  2.5^{+  0.7}_{-  0.9}$ & $135.4^{+  1.2}_{-  1.2}$ \\[2pt]
2M1626-2527 & $0.535^{+0.006}_{-0.007}$ & $-1.07^{+0.49}_{-0.87}$ & $<  4.1$ & $< 25.8$ & $ 1.3^{+ 0.1}_{- 0.1}$ & $   57^{+    5}_{-    4}$ & $0.20^{+0.02}_{-0.02}$ & $ 0.137^{+ 0.010}_{- 0.010}$ & $4.150^{+0.005}_{-0.006}$ & $54.0^{+ 0.5}_{- 0.4}$ & $ 95.5^{+  0.2}_{-  0.2}$ & $138.2^{+  1.4}_{-  1.3}$
\enddata
\end{deluxetable*}

The vertical structure of the disk is governed by hydrostatic equilibrium,
\begin{equation}
\frac{\partial \log \rho}{\partial z} = -\left[\left(\frac{G\,M_*\,z}{(r^2 + z^2)^{3/2}}\right)\left(\frac{\mu m_H}{k_b T_g}\right) + \frac{\partial \log T_g}{\partial z} \right],
\end{equation}
where $\rho$ is the gas density, $M_*$ is the stellar mass, and $T_g$ is the gas temperature. Moreover, $\mu = 2.37$ is the mean molecular weight of the gas, appropriate for a molecular gas with Solar metallicity, and $z$ is the distance in the vertical direction in cylindrical coordinates. For simplicity, we assume that the disk is vertically isothermal, with a radial temperature dependence of 
\begin{equation}
T_g(r) = T_0 \, \left(\frac{r}{1 \, \mathrm{au}}\right)^{-q}
\end{equation}
where $T_0$ is the temperature at 1 au and $q$ is the temperature power law exponent. Under these assumptions, solving the equation of hydrostatic equilibrium finds that the density structure is given by,
\begin{equation}
\rho\,(r,z) = \frac{\Sigma(r)}{\sqrt{2 \pi} \, h(r)} \, \exp\left[-\frac{1}{2}\left(\frac{z}{h(r)}\right)^2\right],
\end{equation}
where $h(r)$ is the scale height of the gas. The scale height from hydrostatic equilibrium is found to be,
\begin{equation}
h(r) = \left[\frac{k_b \, r^3 \, T_g(r)}{G \, M_* \, \mu \, m_H}\right]^{1/2}.
\end{equation}
Finally, the number density of the CO gas can be derived from,
\begin{equation}
n_{\rm CO}(r,z) = X_{\rm CO} \, \frac{\rho(r, z)}{\mu m_H},
\end{equation}
where $X_{\rm CO}$ is the abundance of CO relative to H$_2$. We typically assume that $X_{\rm CO} = 1.0\times10^{-4}$.

We also assume that the disk velocity structure is determined by Keplerian rotation, with an azimuthal velocity of,
\begin{equation}
v_{\phi}(r) = \sqrt{\frac{G \, M_*}{r}}.
\end{equation}
The velocities in the vertical and radial directions are assumed to be zero. We do, however, include microturbulent line broadening ($\xi$, in units of km s$^{-1}$) as a parameter in our fit.

This density, temperature, and velocity structure can be input into the \texttt{RADMC-3D} radiative transfer code \citep{Dullemond2012} to generate synthetic channel maps for a given set of model parameters. We assume local thermodynamic equilibrium to generate images, which is appropriate for disks as their densities are much higher than the critical density for CO of $\sim10^3$ cm$^{-3}$. We also leave a number of viewing orientation parameters as free parameters in our fit: inclination ($i$), position angle ($P.A.$), centroid ($x_0$, $y_0$), systemic velocity ($v_{sys}$), and source distance ($d_{pc}$). The synthetic channel maps generated by raytracing in \texttt{RADMC-3D} are Fourier transformed to produce synthetic visibilities, sampled in the $uv$-plane at the same baselines as our datasets, using the GALARIO code \citep{Tazzari2017,Tazzari2018}.

In all, our model includes 14 free parameters: $M_{*}$, $M_{disk}$, $R_{in}$, $R_{disk}$, $\gamma$, $T_0$, $q$, $\xi$, $v_{sys}$, $i$, $d_{pc}$, $P.A.$, $x_0$, and $y_0$. In practice, however, our modeling code is flexible, and allows additional model features and parameters to be included, for example gaps in the gas density distribution, but here we restrict the model to these 14. We provide further details on additional parameters in Appendix \ref{section:model_parameters}. 

We fit the model channel map visibilities generated by RADMC-3D and GALARIO to our data in the $uv$-plane using the MCMC fitting code \texttt{emcee} \citep{ForemanMackey2013}. For brevity, and because \texttt{emcee} has been broadly used by the community, we opt to provide further details on the setup of our MCMC runs, and details of the convergence in Appendix \ref{section:mcmc_fitting}. The best-fit values are determined by discarding the burn-in MCMC steps and calculating the median of the sample of post-burn-in walker positions for each parameter. We also estimate the uncertainty on the measured parameters by considering the range around the median value that contains 68\% of the post-burn-in walker positions. We provide further details on this process, as well as posterior probability distribution function triangle plots and plots of the walkers' steps in each parameter, in Appendix \ref{section:mcmc_fitting}.

\subsection{Masses and Stellar Parameters from Evolutionary Tracks}
\label{section:evolutionary_track_masses}

Although masses have been previously estimated for 2M1626--2527, CT Cha, and DH Tau using evolutionary models, they were done across separate studies that do not always use uniform methodologies. As such, we find it worthwhile to repeat those measurements in a uniform way, following the method outlined in \cite{Andrews2013} and \cite{Czekala2015}. 

\begin{figure}[t]
\centering
\figurenum{5}
\includegraphics[width=3.3in]{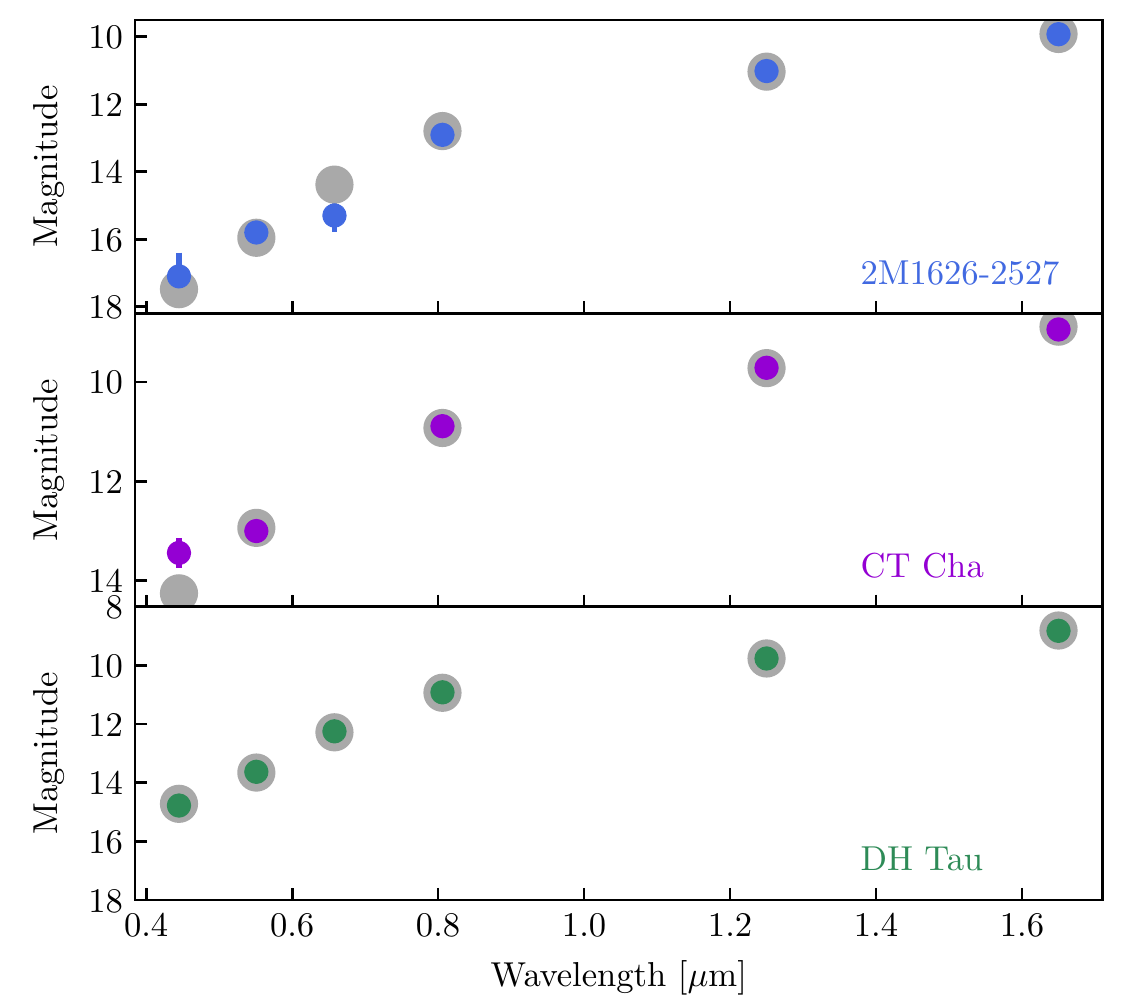}
\caption{Optical and near-infrared photometry for our sources, with the best fit BT-Settl model for each source shown in gray.}
\label{fig:photometry_models}
\end{figure}

To begin, we derived the effective temperature ($T_{eff}$) and bolometric luminosity ($L_*$) by fitting optical and near-infrared photometry collected from the literature \citep{Vrba1989,Epchtein1999,Briceno2002,Ducati2002,Cutri2003,Monet2003,Zacharias2003,Torres2006,Robberto2012,Muinos2014,Henden2016,Bowler2017} with the BT-Settl atmospheric models \citep{Allard2011} using the {\it emcee} package \citep{ForemanMackey2013}. For any set of input parameters, the model grid was interpolated to match $T_{eff}$ and surface gravity (log $g$). The overall brightness was scaled by the distance to the source ($d_{pc}$) and the radius of the star ($R_*$), and then the model was reddened by some amount ($A_V$) using the extinction curves from \cite{Weingartner2001} with $R_V = 5.5$. We assumed a Gaussian prior on $d_{pc}$ derived from $Gaia$ parallaxes (see Appendix \ref{section:mcmc_fitting}). Similarly we assumed a Gaussian prior on $T_{eff}$, using temperatures collected from the literature \citep{Herczeg2014,Bowler2017,Manara2016a} with a standard deviation of one spectral sub-class (150 K). Finally, the stellar luminosity ($L_*$) was calculated from $T_{eff}$ and $R_*$. 

For each post-burn in MCMC sample, we mapped $T_{eff}$ and $L_*$ on to evolutionary tracks to build up a probabilistic sample of stellar masses ($M_*$) and ages ($\tau_*$). The stellar mass and age, and the associated uncertainties, were estimated from these samples using the median and range around the median containing 68\% of the samples. This process was repeated for each of the sets of evolutionary tracks that we considered: the \citet{Baraffe2015} models, the PARSEC models \citep{Bressan2012} and subsequent updates to low-mass ($<0.75~M_\sun$) stars \citep{Chen2014}, as well as both the non-magnetic and magnetic tracks from \cite{Feiden2016}.

\begin{figure}[t]
\centering
\figurenum{6}
\includegraphics[width=3.3in]{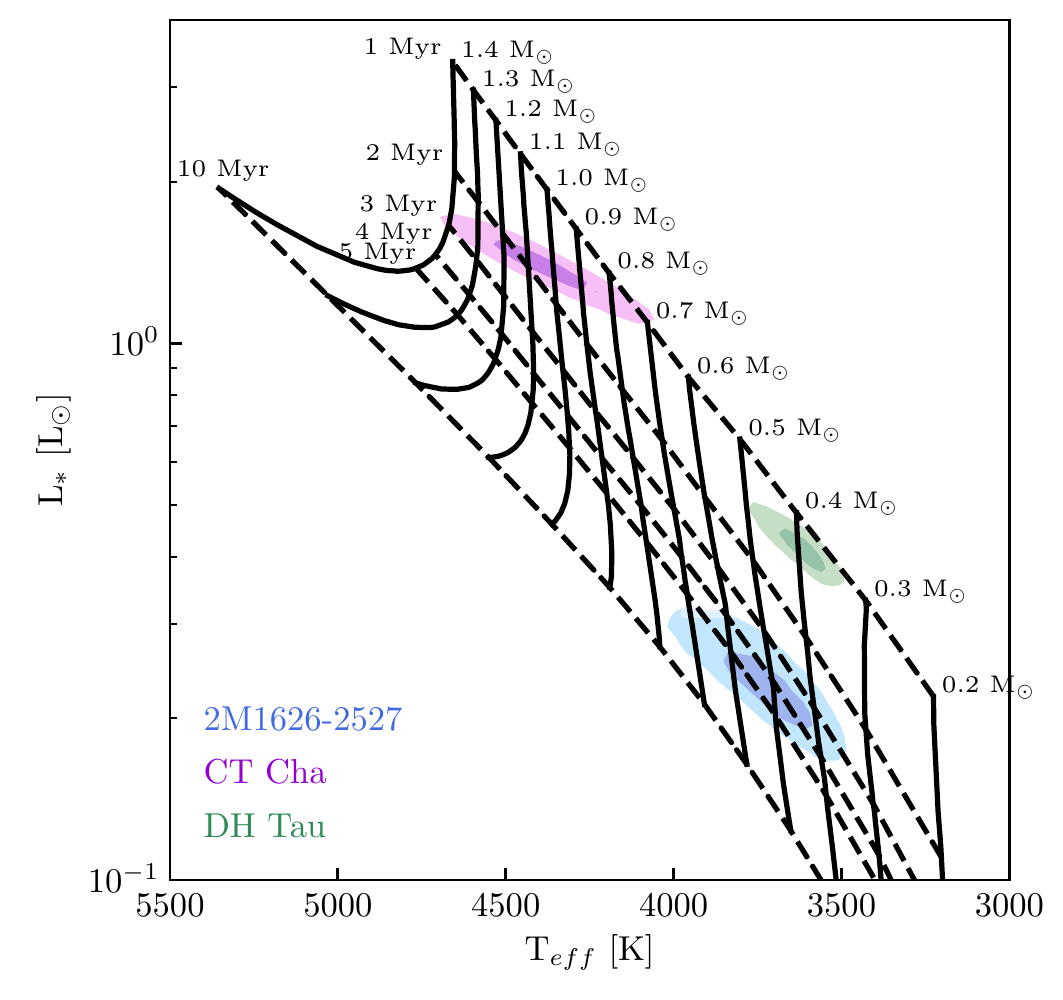}
\caption{Pre-main sequence evolutionary tracks from \citet{Baraffe2015} with the two dimensional posterior probability density functions's of temperature and luminosity derived from our stellar photosphere modeling. By comparing these posterior distributions with evolutionary tracks, masses and ages can be inferred.}
\label{fig:evolutionary_tracks}
\end{figure}

%The stellar mass and age were thus estimated as the median of the samples, with uncertainties containing 68\% of the samples. We find that the inferred stellar mass is highly model-dependent: $\sim$0.4--0.8 $M_\sun$ for 2M1626--2527, $\sim$0.8--1.4 $M_\sun$ for CT Cha, and $\sim$0.4--0.8 $M_\sun$ for DH Tau. Significant systematics apparently exist between evolutionary models. Table \ref{table:stellar_fits} lists the masses and other stellar parameters derived from models. 

%\subsection{Masses and Stellar Parameters from ALMA Measurements}
%\label{section:ALMA_masses}

\vspace{10pt}

\section{Results}
\label{section:results}

\begin{figure*}[t]
\centering
\figurenum{7}
\includegraphics[width=7in]{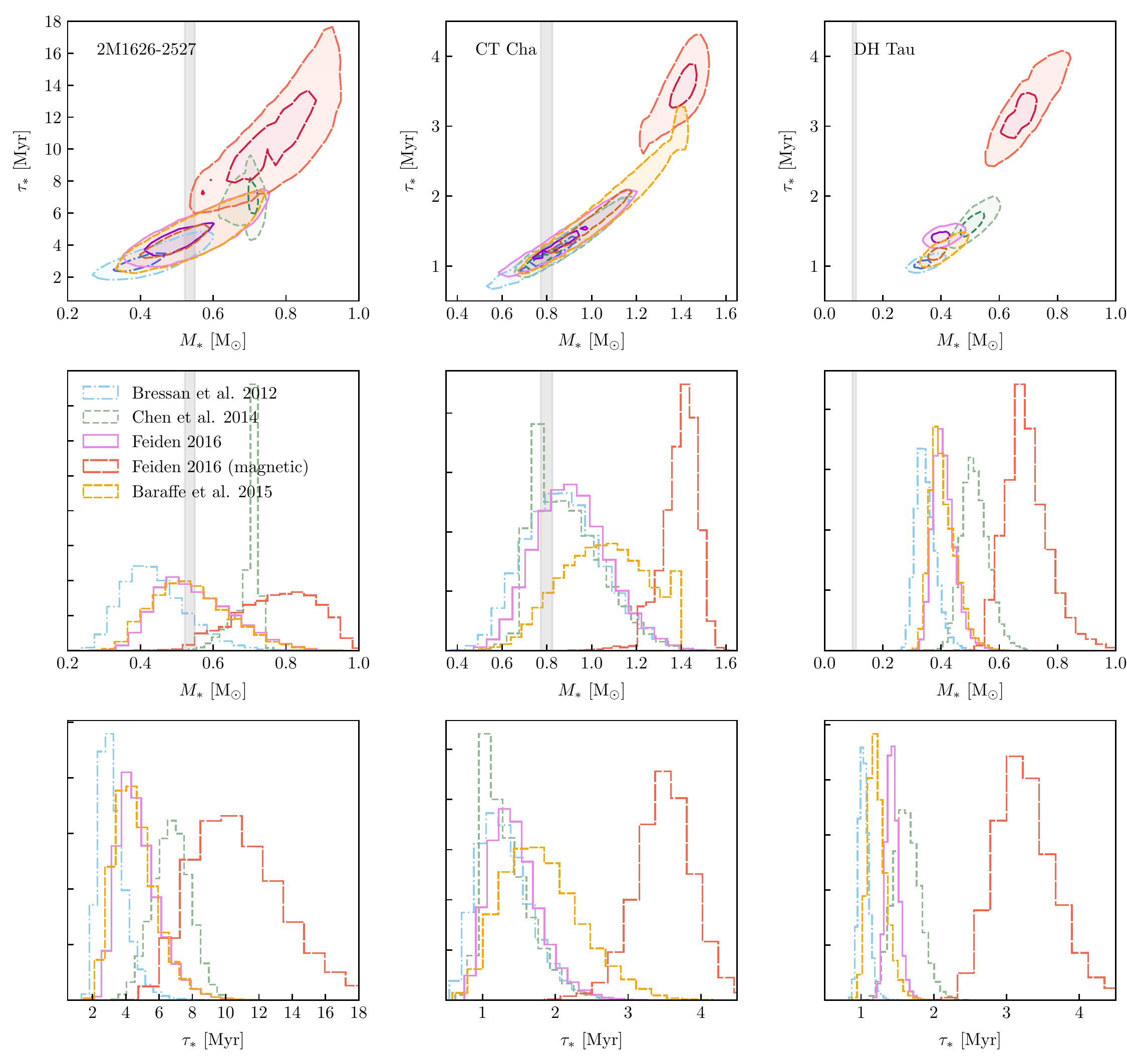}
\caption{Histograms of the posterior probability density functions of stellar mass ($M_*$) and age ($\tau_*$) for 2M1626--2527, CT Cha, and DH Tau. Different colors correspond to different evolutionary tracks, with the key shown in the center-left plot. The $95\%$ confidence mass range derived from our Keplerian disk model fitting is shown as a grey band.}
\label{fig:mass_histograms}
\end{figure*}

We show the best fit Keplerian disk models, as determined by our spectral line fitting procedure, to 2M1626--2527, CT Cha, and DH Tau in Figures \ref{fig:short_model_ROXs12}, \ref{fig:short_model_CTCha}, and \ref{fig:short_model_DHTau}. In these figures we show only a subset of channels to demonstrate that our models provide excellent fits to the data, but the comparison of the models to the full channel map data can be found in Appendix \ref{section:channel_maps}. For all three sources the models provide good fits to the data, with no (or few) significant, $>3\sigma$, residuals. The best fit parameters found from these fits are listed in Table \ref{table:best_fits}. All three disks are small, with characteristic radii of $\lesssim15$ au, although due to the exponential tapering our disk models do contain a small amount of mass at larger radii.

Table \ref{table:stellar_fits} lists the masses and other stellar parameters derived from the stellar photosphere models, and Figure \ref{fig:photometry_models} shows the best fit models compared with the data. Figure \ref{fig:evolutionary_tracks} demonstrates the process of mapping $T_{eff}$ and $L_*$ on to the evolutionary tracks from \cite{Baraffe2015} to estimate $M_*$ and $\tau_*$. The posterior distributions for stellar mass and age for each set of evolutionary tracks are shown in Figure \ref{fig:mass_histograms}, with a gray line representing the mass we derive from our spectral line fits. 

We find that the dynamical masses we measure for 2M1626--2527 and CT Cha are in reasonable agreement with masses inferred from evolutionary tracks, although we discuss the agreement for each source in more detail below. The magnetic tracks from \citet{Feiden2016}, however, seem to universally over-estimate the masses for all three sources.

\subsection{2M1626--2527}
\label{subsection:2M1626--2527}

We find from our spectral line modeling that 2M1626--2527 has a dynamical mass of $0.535^{+0.006}_{-0.007}$ $M_\sun$, in very good agreement with the mass estimated by the \citet{Baraffe2015} and non-magnetic \citet{Feiden2016} evolutionary tracks. Interestingly, though, the original PARSEC models by \citet{Bressan2012} under-predict the mass by $\sim1\sigma$ while the revision to the PARSEC models by \citet{Chen2014}, intended to improve the tracks for stellar masses $<0.75$ $M_\sun$, over-predict 2M1626--2527's mass by $\gtrsim$ 3$\sigma$. Our mass measurement is also consistent with the value of 0.5 $\pm$ 0.1~$M_\sun$ derived by \cite{Bowler2017}, who also used the \citet{Baraffe2015} tracks. 

We also find that the spin axis of 2M1626--2527 is probably not orthogonal to the disk plane (disk inclination of 54\degr~versus stellar inclination of 9\degr--27\degr~found by \cite{Bowler2017}). A similar misalignment of disk and star was seen in our previous study of GQ Lup \citep{Wu2017a}. %We note that the systemic velocity of 4.15 km s$^{-1}$ is about 2--3 km s$^{-1}$ lower than the measured radial velocity of the star \citep{Bowler2017}, which could arise from stellar jitters due to magnetic activities, or may hint at an unseen companion.

\subsection{CT Cha}

CT Cha has a stellar mass of $0.796^{+0.015}_{-0.014}$ $M_{\odot}$, which, contrary to 2M1626--2527, is in very good agreement with both sets of PARSEC evolutionary tracks. The non-magnetic \citet{Feiden2016} tracks are also in reasonable, $\lesssim0.5\sigma$, agreement, while the \citet{Baraffe2015} evolutionary tracks predict a mass that is $\sim1\sigma$ too high. Furthermore, the mass is in reasonable ($\lesssim0.5\sigma$) agreement with the mass of $0.87^{+0.23}_{-0.18}~M_{\odot}$ estimated by \citet{Pascucci2016} using a combination of the \citet{Baraffe2015} and \citet{Feiden2016} tracks. 

%The disk mass we find for CT Cha is also quite high, at $M_{disk} > 0.08$ $M_{\odot}$. The lower limit is likely an indication that the disk is significantly optically thick. The disk mass values measured here are also likely more uncertain than the statistical uncertainties indicate, as we have made relatively simplistic assumptions about the abundance of CO. Other studies have shown that the CO abundance can be significantly different from the commonly assumed value of $\sim10^{-4}$ \citep[e.g.,][]{Schwarz2016,Ansdell2016,Miotello2017}. A model that employs more detailed chemistry is likely needed to  truly understand the gas masses of these disks.

\begin{deluxetable}{lcccccc}
\tablecaption{Stellar Photosphere Model Parameters \& Stellar Parameters Inferred from Evolutionary Tracks}
\tablenum{3}
\tabletypesize{\small}
\label{table:stellar_fits}
\tablehead{\colhead{Parameters} & \colhead{2M1626-2527} & \colhead{CT Cha} & \colhead{DH Tau}}
\startdata
$T_{eff}$ [K] & $3746^{+139}_{-122}$ & $4402^{+151}_{-166}$ & $3628^{+ 81}_{- 65}$ \\[2pt]
$\log\,g$ & $3.6^{+0.6}_{-0.7}$ & $4.5^{+0.7}_{-1.2}$ & $3.1^{+0.4}_{-0.4}$ \\[2pt]
$R_*$ [$R_{\odot}$] & $1.17^{+0.06}_{-0.04}$ & $2.06^{+0.05}_{-0.05}$ & $1.66^{+0.04}_{-0.03}$ \\[2pt]
$d_{pc}$ [pc] & $138.3^{+  1.4}_{-  1.4}$ & $191.8^{+  0.8}_{-  0.8}$ & $135.4^{+  1.2}_{-  1.3}$ \\[2pt]
$A_V$ & $2.5^{+0.6}_{-0.6}$ & $1.6^{+0.3}_{-0.3}$ & $0.5^{+0.3}_{-0.3}$ \\[2pt]
$L_*$ [$L_{\odot}$] & $0.24^{+0.05}_{-0.04}$ & $1.41^{+0.17}_{-0.16}$ & $0.42^{+0.04}_{-0.04}$ \\[2pt]
\tableline
\multicolumn{4}{c}{$M_*$ [M$_{\odot}$]} \\[2pt]
\tableline
Baraffe et al. 2015 & $0.53^{+0.12}_{-0.10}$ & $1.06^{+0.20}_{-0.20}$ & $0.40^{+0.05}_{-0.03}$ \\[2pt]
Bressan et al. 2012 & $0.43^{+0.10}_{-0.08}$ & $0.87^{+0.18}_{-0.17}$ & $0.34^{+0.04}_{-0.03}$ \\[2pt]
Chen et al. 2014 & $0.71^{+0.01}_{-0.04}$ & $0.87^{+0.18}_{-0.13}$ & $0.51^{+0.05}_{-0.04}$ \\[2pt]
Feiden 2016 & $0.54^{+0.12}_{-0.09}$ & $0.90^{+0.17}_{-0.15}$ & $0.41^{+0.04}_{-0.03}$ \\[2pt]
Feiden 2016 (magnetic) & $0.78^{+0.11}_{-0.12}$ & $1.41^{+0.06}_{-0.09}$ & $0.69^{+0.08}_{-0.06}$ \\[2pt]
\tableline
\multicolumn{4}{c}{Age [Myr]} \\[2pt]
\tableline
Baraffe et al. 2015 & $4.41^{+1.43}_{-1.06}$ & $1.77^{+0.60}_{-0.49}$ & $1.21^{+0.14}_{-0.10}$ \\[2pt]
Bressan et al. 2012 & $3.07^{+0.92}_{-0.62}$ & $1.26^{+0.41}_{-0.32}$ & $1.04^{+0.07}_{-0.06}$ \\[2pt]
Chen et al. 2014 & $6.76^{+1.10}_{-1.14}$ & $1.26^{+0.41}_{-0.23}$ & $1.62^{+0.19}_{-0.17}$ \\[2pt]
Feiden 2016 & $4.56^{+1.41}_{-0.95}$ & $1.41^{+0.38}_{-0.30}$ & $1.43^{+0.09}_{-0.08}$ \\[2pt]
Feiden 2016 (magnetic) & $10.56^{+3.15}_{-2.50}$ & $3.51^{+0.38}_{-0.39}$ & $3.23^{+0.47}_{-0.37}$ \\[2pt]
\enddata
\end{deluxetable}

\subsection{DH Tau}
\label{subsection:DHTau}

We find that DH Tau has a mass of $0.101^{+0.004}_{-0.003}$ $M_{\odot}$, significantly lower than 0.3--0.5 $M_{\odot}$ we derive from evolutionary track models. It is also significantly lower than the 0.2--0.5 $M_{\odot}$ estimated by \citet{Hartigan1994}, \citet{White2001}, and \citet{Andrews2013}. However, it is possible that our modeling is affected by the strong molecular cloud absorption obscuring much of the low velocity emission. In order to test whether foreground extinction could be affecting our mass measurement for DH Tau, we add a simple model for the extinction to our fit. We assume that each channel is extincted by an amount of $\exp{(-\tau_{\nu})}$. We assume that the optical depth, $\tau_{\nu}$, is Gaussian in shape with some central velocity, $v_{0,ext}$, width $\sigma_{ext}$, and peak value, $\tau_0$. We otherwise use the same model described above and fit the model to the data using the same methods, but with three additional parameters.

We find that this new model does increase the stellar mass to $0.141^{+0.016}_{-0.011}$ $M_\sun$. The difference in mass can be attributed to the lower inclination found by this model, of $38.0^{+ 2.4}_{- 2.5}$ degrees. The best fit values of the new parameters are $v_{0,ext} = 6.629^{+0.016}_{-0.018}$ km s$^{-1}$, $\sigma_{ext} = 0.389^{+0.027}_{-0.026}$ km s$^{-1}$, and $\tau_0 = 6.15^{+1.71}_{-1.01}$, and all other parameters are consistent with the values we find for our base model. Although this simple model cannot fully reconcile the mass discrepancy between our measurement and the evolutionary tracks, it is possible that a more complicated extinction profile could increase the inferred stellar mass even more.

\vspace{10pt}

\section{Discussion}
\label{section:discussion}

The dynamical masses we measure from our Keplerian disk fits demonstrates the power of combining ALMA spectral line maps with precise distance measurements with $Gaia$. Previous Keplerian disk dynamical mass measurements exhibited extreme degeneracies between stellar mass and source distance \citep[e.g.,][]{Czekala2015}. As uncertainties in the source distance could be as large as 10--20\% or more, they translated directly into 10--20\% uncertainty on the measured stellar mass. Here, using precise estimates of distances from $Gaia$, our stellar mass estimates have uncertainties of only 2-4\%. 

As the uncertainty on $Gaia$ distances is $\lesssim1\%$, it seems likely that these uncertainties are no longer dominated by the distance uncertainty. Indeed, the posterior probability density function plots shown in Appendix \ref{section:mcmc_fitting} show little degeneracy between stellar mass and source distance for DH Tau and CT Cha, suggesting that the measurement is limited primarily by the quality of the ALMA data. The ALMA data for 2M1626--2527 are good enough that the $Gaia$ distance uncertainty does affect the mass measurement, noticeable as slight a degeneracy between mass and distance, but the $Gaia$ uncertainty is only $\sim1\%$ so it does not prevent us from making a high precision measurement of the stellar mass.

Interestingly, the comparison of our dynamical mass measurements with evolutionary track mass estimates shows a wide range in agreement. The PARSEC evolutionary tracks \citep{Bressan2012,Chen2014} do the best job of reproducing the highest mass source in our sample, CT Cha, but fail to correctly reproduce the mass of 2M1626--2527. Conversely, the \citet{Baraffe2015} evolutionary tracks do an excellent job of reproducing the mass of 2M1626--2527, but overestimate the mass of CT Cha. The non-magnetic tracks from \citet{Feiden2016} may provide the best balance of matching the masses of both CT Cha and 2M1626--2527, but do not match both perfectly. It is clear, though, that the magnetic tracks from \citet{Feiden2016} do a poor job of reproducing the masses of both sources.

The case of DH Tau is particularly interesting because the mass we measure is discrepant from evolutionary track estimates, which are uniformly ($>4\sigma$) too high. As we demonstrated earlier, foreground extinction could affect our stellar mass measurement, although our simple model for extinction only increases the stellar mass to $\sim$0.14 $M_\sun$ and does not reconcile with evolutionary tracks. Of course, a more complicated extinction profile may increase the stellar mass by more. If the mass estimate is correct, however, then a relatively major shift in the evolutionary tracks for young, low-mass objects is needed. Higher signal-to-noise data may help to better understand the effects of cloud contamination on this source and better constrain its mass, as might more optically thin gas tracers that better see through the foreground cloud. For now, however, it remains unclear whether the discrepancy is a result of the limitations of the data, or whether there are real problems with the evolutionary tracks.

Moving forward, however, a much larger sample of pre-main sequence stars is needed to understand whether the trends suggested by our targets are more generally true, and thereby truly place constraints on evolutionary tracks. Our results here demonstrate that with relatively modest observations ($\sim$15 minutes on source; $0\farcs2-0\farcs3$ resolution; 0.4 km s$^{-1}$ channels) we can robustly measure stellar masses as low as $\sim$0.1 $M_\sun$. Moreover, with channel widths as small as 0.03--0.05 km s$^{-1}$ for CO (2--1) or CO (3--2), it should be possible to directly measure masses for young sub-stellar and planetary mass objects. Building a sample with several tens of pre-main sequence sources could therefore reasonably be done.

\section{Conclusions}
\label{section:conclusion}

We have used our Keplerian disk radiative transfer code \texttt{psdpy} to fit ALMA $^{12}$CO (2--1) channel maps for three T Tauri stars 2M1626--2527, CT Cha, and DH Tau. With new, precise distance measurements for these sources from $Gaia$, we are able to make high precision ($\sim$2--4\%) direct measurements of their masses. The mass uncertainty is no longer dominated by distance as in previous studies, but instead by the quality of ALMA data. We find that 2M1626--2527 has a mass of $0.535^{+0.006}_{-0.007}$ M$_{\odot}$, CT Cha has a mass of $0.796^{+0.015}_{-0.014}$ M$_{\odot}$, and DH Tau has a mass of $0.101^{+0.004}_{-0.003}$ M$_{\odot}$. Comparing with the stellar masses estimated using multiple sets of evolutionary tracks, we find that both 2M1626--2527 and CT Cha are in reasonable agreement with most models ($<$ 2$\sigma$), although potentially significant differences remain for most sets of tracks. The large discrepancy for DH Tau may be caused by a significant local cloud obscuration, which completely absorbs the redshifted disk emission, although simple models of the foreground extinction only increase the mass to $0.141^{+0.016}_{-0.011}$ $M_\sun$.

Looking towards the future, the synergy of ALMA, $Gaia$, and detailed radiative transfer disk modeling will enable precise mass measurements of a large number of pre-main sequence stars, and even for substellar objects, and ultimately place stringent constraints on evolutionary models.

\software{pdspy \citep{Sheehan2018b}, CASA \citep{McMullin2007}, RADMC-3D \citep{Dullemond2012}, emcee \citep{ForemanMackey2013}, matplotlib \citep{Hunter2007}, corner \citep{ForemanMackey2016}, GALARIO \citep{Tazzari2018}}

\acknowledgements

We would like the thank the referee for a number of comments that helped to significantly improve this work. This paper makes use of the following ALMA data: ADS/JAO.ALMA\#2015.1.00773.S. ALMA is a partnership of ESO (representing its member states), NSF (USA) and NINS (Japan), together with NRC (Canada), NSC and ASIAA (Taiwan), and KASI (Republic of Korea), in cooperation with the Republic of Chile. The Joint ALMA Observatory is operated by ESO, AUI/NRAO, and NAOJ. The National Radio Astronomy Observatory is a facility of the National Science Foundation operated under cooperative agreement by Associated Universities, Inc. Y.-L. W. is grateful to the support from the Heising-Simons Foundation. J.T. would like to acknowledge support from the Homer L. Dodge Endowed Chair. This material is based upon work supported by the National Science Foundation Graduate Research Fellowship under Grant No. 2012115762. This work was supported by NSF AAG grant 1311910. The results reported herein benefitted from collaborations and/or information exchange within NASA's Nexus for Exoplanet System Science (NExSS) research coordination network sponsored by NASA's Science Mission Directorate. The computing for this project was performed at the OU Supercomputing Center for Education \& Research (OSCER) at the University of Oklahoma (OU).

\bibliography{ms.bib}

\begin{thebibliography}{}
\expandafter\ifx\csname natexlab\endcsname\relax\def\natexlab#1{#1}\fi

\bibitem[{{Allard} {et~al.}(2011){Allard}, {Homeier}, \&
  {Freytag}}]{Allard2011}
{Allard}, F., {Homeier}, D., \& {Freytag}, B. 2011, in Astronomical Society of
  the Pacific Conference Series, Vol. 448, 16th Cambridge Workshop on Cool
  Stars, Stellar Systems, and the Sun, ed. C.~{Johns-Krull}, M.~K. {Browning},
  \& A.~A. {West}, 91

\bibitem[{{Andrews} {et~al.}(2013){Andrews}, {Rosenfeld}, {Kraus}, \&
  {Wilner}}]{Andrews2013}
{Andrews}, S.~M., {Rosenfeld}, K.~A., {Kraus}, A.~L., \& {Wilner}, D.~J. 2013,
  \apj, 771, 129

\bibitem[{{Ansdell} {et~al.}(2016){Ansdell}, {Williams}, {van der Marel},
  {Carpenter}, {Guidi}, {Hogerheijde}, {Mathews}, {Manara}, {Miotello},
  {Natta}, {Oliveira}, {Tazzari}, {Testi}, {van Dishoeck}, \& {van
  Terwisga}}]{Ansdell2016}
{Ansdell}, M., {Williams}, J.~P., {van der Marel}, N., {et~al.} 2016, \apj,
  828, 46

\bibitem[{{Ansdell} {et~al.}(2018){Ansdell}, {Williams}, {Trapman}, {van
  Terwisga}, {Facchini}, {Manara}, {van der Marel}, {Miotello}, {Tazzari},
  {Hogerheijde}, {Guidi}, {Testi}, \& {van Dishoeck}}]{Ansdell2018}
{Ansdell}, M., {Williams}, J.~P., {Trapman}, L., {et~al.} 2018, \apj, 859, 21

\bibitem[{{Baraffe} {et~al.}(2015){Baraffe}, {Homeier}, {Allard}, \&
  {Chabrier}}]{Baraffe2015}
{Baraffe}, I., {Homeier}, D., {Allard}, F., \& {Chabrier}, G. 2015, \aap, 577,
  A42

\bibitem[{{Barenfeld} {et~al.}(2016){Barenfeld}, {Carpenter}, {Ricci}, \&
  {Isella}}]{Barenfeld2016}
{Barenfeld}, S.~A., {Carpenter}, J.~M., {Ricci}, L., \& {Isella}, A. 2016,
  \apj, 827, 142

\bibitem[{{Bell}(2016)}]{Bell2016}
{Bell}, C. P.~M. 2016, in 19th Cambridge Workshop on Cool Stars, Stellar
  Systems, and the Sun (CS19), 102

\bibitem[{{Bowler} {et~al.}(2017){Bowler}, {Kraus}, {Bryan}, {Knutson},
  {Brogi}, {Rizzuto}, {Mace}, {Vanderburg}, {Liu}, {Hillenbrand}, \&
  {Cieza}}]{Bowler2017}
{Bowler}, B.~P., {Kraus}, A.~L., {Bryan}, M.~L., {et~al.} 2017, \aj, 154, 165

\bibitem[{{Bressan} {et~al.}(2012){Bressan}, {Marigo}, {Girardi}, {Salasnich},
  {Dal Cero}, {Rubele}, \& {Nanni}}]{Bressan2012}
{Bressan}, A., {Marigo}, P., {Girardi}, L., {et~al.} 2012, \mnras, 427, 127

\bibitem[{{Brice{\~n}o} {et~al.}(2002){Brice{\~n}o}, {Luhman}, {Hartmann},
  {Stauffer}, \& {Kirkpatrick}}]{Briceno2002}
{Brice{\~n}o}, C., {Luhman}, K.~L., {Hartmann}, L., {Stauffer}, J.~R., \&
  {Kirkpatrick}, J.~D. 2002, \apj, 580, 317

\bibitem[{{Chabrier}(2003)}]{Chabrier2003}
{Chabrier}, G. 2003, \pasp, 115, 763

\bibitem[{{Chen} {et~al.}(2014){Chen}, {Girardi}, {Bressan}, {Marigo},
  {Barbieri}, \& {Kong}}]{Chen2014}
{Chen}, Y., {Girardi}, L., {Bressan}, A., {et~al.} 2014, \mnras, 444, 2525

\bibitem[{{Cutri} {et~al.}(2003){Cutri}, {Skrutskie}, {van Dyk}, {Beichman},
  {Carpenter}, {Chester}, {Cambresy}, {Evans}, {Fowler}, {Gizis}, {Howard},
  {Huchra}, {Jarrett}, {Kopan}, {Kirkpatrick}, {Light}, {Marsh}, {McCallon},
  {Schneider}, {Stiening}, {Sykes}, {Weinberg}, {Wheaton}, {Wheelock}, \&
  {Zacarias}}]{Cutri2003}
{Cutri}, R.~M., {Skrutskie}, M.~F., {van Dyk}, S., {et~al.} 2003, VizieR Online
  Data Catalog, 2246

\bibitem[{{Czekala} {et~al.}(2015){Czekala}, {Andrews}, {Jensen}, {Stassun},
  {Torres}, \& {Wilner}}]{Czekala2015}
{Czekala}, I., {Andrews}, S.~M., {Jensen}, E.~L.~N., {et~al.} 2015, \apj, 806,
  154

\bibitem[{{Czekala} {et~al.}(2016){Czekala}, {Andrews}, {Torres}, {Jensen},
  {Stassun}, {Wilner}, \& {Latham}}]{Czekala2016}
{Czekala}, I., {Andrews}, S.~M., {Torres}, G., {et~al.} 2016, \apj, 818, 156

\bibitem[{{Czekala} {et~al.}(2017){Czekala}, {Andrews}, {Torres}, {Rodriguez},
  {Jensen}, {Stassun}, {Latham}, {Wilner}, {Gully-Santiago}, {Grankin}, {Lund},
  {Kuhn}, {Stevens}, {Siverd}, {James}, {Gaudi}, {Shappee}, \&
  {Holoien}}]{Czekala2017}
---. 2017, \apj, 851, 132

\bibitem[{{Ducati}(2002)}]{Ducati2002}
{Ducati}, J.~R. 2002, VizieR Online Data Catalog

\bibitem[{{Dullemond}(2012)}]{Dullemond2012}
{Dullemond}, C.~P. 2012, {RADMC-3D: A multi-purpose radiative transfer tool},
  Astrophysics Source Code Library, , , ascl:1202.015

\bibitem[{{Dupuy} {et~al.}(2016){Dupuy}, {Forbrich}, {Rizzuto}, {Mann},
  {Aller}, {Liu}, {Kraus}, \& {Berger}}]{Dupuy2016}
{Dupuy}, T.~J., {Forbrich}, J., {Rizzuto}, A., {et~al.} 2016, \apj, 827, 23

\bibitem[{{Dutrey} {et~al.}(1994){Dutrey}, {Guilloteau}, \&
  {Simon}}]{Dutrey1994}
{Dutrey}, A., {Guilloteau}, S., \& {Simon}, M. 1994, \aap, 286, 149

\bibitem[{{Dutrey} {et~al.}(2003){Dutrey}, {Guilloteau}, \&
  {Simon}}]{Dutrey2003}
---. 2003, \aap, 402, 1003

\bibitem[{{Eisner} {et~al.}(2018){Eisner}, {Arce}, {Ballering}, {Bally},
  {Andrews}, {Boyden}, {Di Francesco}, {Fang}, {Johnstone}, {Kim}, {Mann},
  {Matthews}, {Pascucci}, {Ricci}, {Sheehan}, \& {Williams}}]{Eisner2018}
{Eisner}, J.~A., {Arce}, H.~G., {Ballering}, N.~P., {et~al.} 2018, \apj, 860,
  77

\bibitem[{{Epchtein} {et~al.}(1999){Epchtein}, {Deul}, {Derriere},
  {Borsenberger}, {Egret}, {Simon}, {Alard}, {Bal{\'a}zs}, {de Batz}, {Cioni},
  {Copet}, {Dennefeld}, {Forveille}, {Fouqu{\'e}}, {Garz{\'o}n}, {Habing},
  {Holl}, {Hron}, {Kimeswenger}, {Lacombe}, {Le Bertre}, {Loup}, {Mamon},
  {Omont}, {Paturel}, {Persi}, {Robin}, {Rouan}, {Tiph{\`e}ne}, {Vauglin}, \&
  {Wagner}}]{Epchtein1999}
{Epchtein}, N., {Deul}, E., {Derriere}, S., {et~al.} 1999, \aap, 349, 236

\bibitem[{{Feiden}(2016)}]{Feiden2016}
{Feiden}, G.~A. 2016, \aap, 593, A99

\bibitem[{{Flaherty} {et~al.}(2015){Flaherty}, {Hughes}, {Rosenfeld},
  {Andrews}, {Chiang}, {Simon}, {Kerzner}, \& {Wilner}}]{Flaherty2015}
{Flaherty}, K.~M., {Hughes}, A.~M., {Rosenfeld}, K.~A., {et~al.} 2015, \apj,
  813, 99

\bibitem[{{Flaherty} {et~al.}(2018){Flaherty}, {Hughes}, {Teague}, {Simon},
  {Andrews}, \& {Wilner}}]{Flaherty2018}
{Flaherty}, K.~M., {Hughes}, A.~M., {Teague}, R., {et~al.} 2018, \apj, 856, 117

\bibitem[{{Flaherty} {et~al.}(2017){Flaherty}, {Hughes}, {Rose}, {Simon}, {Qi},
  {Andrews}, {K{\'o}sp{\'a}l}, {Wilner}, {Chiang}, {Armitage}, \&
  {Bai}}]{Flaherty2017}
{Flaherty}, K.~M., {Hughes}, A.~M., {Rose}, S.~C., {et~al.} 2017, \apj, 843,
  150

\bibitem[{Foreman-Mackey(2016)}]{ForemanMackey2016}
Foreman-Mackey, D. 2016, The Journal of Open Source Software, 24,
  doi:10.21105/joss.00024

\bibitem[{{Foreman-Mackey} {et~al.}(2013){Foreman-Mackey}, {Hogg}, {Lang}, \&
  {Goodman}}]{ForemanMackey2013}
{Foreman-Mackey}, D., {Hogg}, D.~W., {Lang}, D., \& {Goodman}, J. 2013, \pasp,
  125, 306

\bibitem[{{Gaia Collaboration} {et~al.}(2018){Gaia Collaboration}, {Brown},
  {Vallenari}, {Prusti}, {de Bruijne}, {Babusiaux}, {Bailer-Jones}, {Biermann},
  {Evans}, {Eyer}, \& et~al.}]{GaiaCollaboration2018}
{Gaia Collaboration}, {Brown}, A.~G.~A., {Vallenari}, A., {et~al.} 2018, \aap,
  616, A1

\bibitem[{{Gennaro} {et~al.}(2012){Gennaro}, {Prada Moroni}, \&
  {Tognelli}}]{Gennaro2012}
{Gennaro}, M., {Prada Moroni}, P.~G., \& {Tognelli}, E. 2012, \mnras, 420, 986

\bibitem[{{Goodman} \& {Weare}(2010)}]{Goodman2010}
{Goodman}, J., \& {Weare}, J. 2010, Communications in Applied Mathematics and
  Computational Science, Vol.~5, No.~1, p.~65-80, 2010, 5, 65

\bibitem[{{Haisch} {et~al.}(2001){Haisch}, {Lada}, \& {Lada}}]{Haisch2001}
{Haisch}, Jr., K.~E., {Lada}, E.~A., \& {Lada}, C.~J. 2001, \apjl, 553, L153

\bibitem[{{Hartigan} {et~al.}(1994){Hartigan}, {Strom}, \&
  {Strom}}]{Hartigan1994}
{Hartigan}, P., {Strom}, K.~M., \& {Strom}, S.~E. 1994, \apj, 427, 961

\bibitem[{{Henden} {et~al.}(2016){Henden}, {Templeton}, {Terrell}, {Smith},
  {Levine}, \& {Welch}}]{Henden2016}
{Henden}, A.~A., {Templeton}, M., {Terrell}, D., {et~al.} 2016, VizieR Online
  Data Catalog, 2336

\bibitem[{{Herczeg} \& {Hillenbrand}(2014)}]{Herczeg2014}
{Herczeg}, G.~J., \& {Hillenbrand}, L.~A. 2014, \apj, 786, 97

\bibitem[{{Hern{\'a}ndez} {et~al.}(2008){Hern{\'a}ndez}, {Hartmann}, {Calvet},
  {Jeffries}, {Gutermuth}, {Muzerolle}, \& {Stauffer}}]{Hernandez2008}
{Hern{\'a}ndez}, J., {Hartmann}, L., {Calvet}, N., {et~al.} 2008, \apj, 686,
  1195

\bibitem[{{Hillenbrand} \& {White}(2004)}]{Hillenbrand2004}
{Hillenbrand}, L.~A., \& {White}, R.~J. 2004, \apj, 604, 741

\bibitem[{Hunter(2007)}]{Hunter2007}
Hunter, J.~D. 2007, Computing In Science \& Engineering, 9, 90

\bibitem[{{Lynden-Bell} \& {Pringle}(1974)}]{LyndenBell1974}
{Lynden-Bell}, D., \& {Pringle}, J.~E. 1974, \mnras, 168, 603

\bibitem[{{MacGregor} {et~al.}(2017){MacGregor}, {Wilner}, {Czekala},
  {Andrews}, {Dai}, {Herczeg}, {Kratter}, {Kraus}, {Ricci}, \&
  {Testi}}]{MacGregor2017}
{MacGregor}, M.~A., {Wilner}, D.~J., {Czekala}, I., {et~al.} 2017, \apj, 835,
  17

\bibitem[{{Manara} {et~al.}(2016){Manara}, {Fedele}, {Herczeg}, \&
  {Teixeira}}]{Manara2016a}
{Manara}, C.~F., {Fedele}, D., {Herczeg}, G.~J., \& {Teixeira}, P.~S. 2016,
  \aap, 585, A136

\bibitem[{{McMullin} {et~al.}(2007){McMullin}, {Waters}, {Schiebel}, {Young},
  \& {Golap}}]{McMullin2007}
{McMullin}, J.~P., {Waters}, B., {Schiebel}, D., {Young}, W., \& {Golap}, K.
  2007, in Astronomical Society of the Pacific Conference Series, Vol. 376,
  Astronomical Data Analysis Software and Systems XVI, ed. R.~A. {Shaw},
  F.~{Hill}, \& D.~J. {Bell}, 127

\bibitem[{{Miotello} {et~al.}(2017){Miotello}, {van Dishoeck}, {Williams},
  {Ansdell}, {Guidi}, {Hogerheijde}, {Manara}, {Tazzari}, {Testi}, {van der
  Marel}, \& {van Terwisga}}]{Miotello2017}
{Miotello}, A., {van Dishoeck}, E.~F., {Williams}, J.~P., {et~al.} 2017, \aap,
  599, A113

\bibitem[{{Monet} {et~al.}(2003){Monet}, {Levine}, {Canzian}, {Ables}, {Bird},
  {Dahn}, {Guetter}, {Harris}, {Henden}, {Leggett}, {Levison}, {Luginbuhl},
  {Martini}, {Monet}, {Munn}, {Pier}, {Rhodes}, {Riepe}, {Sell}, {Stone},
  {Vrba}, {Walker}, {Westerhout}, {Brucato}, {Reid}, {Schoening}, {Hartley},
  {Read}, \& {Tritton}}]{Monet2003}
{Monet}, D.~G., {Levine}, S.~E., {Canzian}, B., {et~al.} 2003, \aj, 125, 984

\bibitem[{{Mui{\~n}os} \& {Evans}(2014)}]{Muinos2014}
{Mui{\~n}os}, J.~L., \& {Evans}, D.~W. 2014, Astronomische Nachrichten, 335,
  367

\bibitem[{{{\"O}berg} {et~al.}(2015){{\"O}berg}, {Furuya}, {Loomis}, {Aikawa},
  {Andrews}, {Qi}, {van Dishoeck}, \& {Wilner}}]{Oberg2015}
{{\"O}berg}, K.~I., {Furuya}, K., {Loomis}, R., {et~al.} 2015, \apj, 810, 112

\bibitem[{{Ortiz-Le{\'o}n} {et~al.}(2017){Ortiz-Le{\'o}n}, {Loinard},
  {Kounkel}, {Dzib}, {Mioduszewski}, {Rodr{\'{\i}}guez}, {Torres},
  {Gonz{\'a}lez-L{\'o}pezlira}, {Pech}, {Rivera}, {Hartmann}, {Boden}, {Evans},
  {Brice{\~n}o}, {Tobin}, {Galli}, \& {Gudehus}}]{OrtizLeon2017}
{Ortiz-Le{\'o}n}, G.~N., {Loinard}, L., {Kounkel}, M.~A., {et~al.} 2017, \apj,
  834, 141

\bibitem[{{Pascucci} {et~al.}(2016){Pascucci}, {Testi}, {Herczeg}, {Long},
  {Manara}, {Hendler}, {Mulders}, {Krijt}, {Ciesla}, {Henning}, {Mohanty},
  {Drabek-Maunder}, {Apai}, {Szucs}, {Sacco}, \& {Olofsson}}]{Pascucci2016}
{Pascucci}, I., {Testi}, L., {Herczeg}, G.~J., {et~al.} 2016, ArXiv e-prints,
  arXiv:1608.03621

\bibitem[{{Ricci} {et~al.}(2017){Ricci}, {Cazzoletti}, {Czekala}, {Andrews},
  {Wilner}, {Sz{\H u}cs}, {Lodato}, {Testi}, {Pascucci}, {Mohanty}, {Apai},
  {Carpenter}, \& {Bowler}}]{Ricci2017}
{Ricci}, L., {Cazzoletti}, P., {Czekala}, I., {et~al.} 2017, \aj, 154, 24

\bibitem[{{Rizzuto} {et~al.}(2016){Rizzuto}, {Ireland}, {Dupuy}, \&
  {Kraus}}]{Rizzuto2016}
{Rizzuto}, A.~C., {Ireland}, M.~J., {Dupuy}, T.~J., \& {Kraus}, A.~L. 2016,
  \apj, 817, 164

\bibitem[{{Robberto} {et~al.}(2012){Robberto}, {Spina}, {Da Rio}, {Apai},
  {Pascucci}, {Ricci}, {Goddi}, {Testi}, {Palla}, \&
  {Bacciotti}}]{Robberto2012}
{Robberto}, M., {Spina}, L., {Da Rio}, N., {et~al.} 2012, \aj, 144, 83

\bibitem[{{Rodet} {et~al.}(2018){Rodet}, {Bonnefoy}, {Durkan}, {Beust},
  {Lagrange}, {Schlieder}, {Janson}, {Grandjean}, {Chauvin}, {Messina},
  {Maire}, {Brandner}, {Girard}, {Delorme}, {Biller}, {Bergfors}, {Lacour},
  {Feldt}, {Henning}, {Boccaletti}, {Le Bouquin}, {Berger}, {Monin}, {Udry},
  {Peretti}, {Segransan}, {Allard}, {Homeier}, {Vigan}, {Langlois},
  {Hagelberg}, {Menard}, {Bazzon}, {Beuzit}, {Delboulbe}, {Desidera},
  {Gratton}, {Lannier}, {Ligi}, {Maurel}, {Mesa}, {Meyer}, {Pavlov}, {Ramos},
  {Rigal}, {Roelfsema}, {Salter}, {Samland}, {Schmidt}, {Stadler}, \&
  {Weber}}]{Rodet2018}
{Rodet}, L., {Bonnefoy}, M., {Durkan}, S., {et~al.} 2018, ArXiv e-prints,
  arXiv:1806.05491

\bibitem[{{Schaefer} {et~al.}(2009){Schaefer}, {Dutrey}, {Guilloteau}, {Simon},
  \& {White}}]{Schaefer2009}
{Schaefer}, G.~H., {Dutrey}, A., {Guilloteau}, S., {Simon}, M., \& {White},
  R.~J. 2009, \apj, 701, 698

\bibitem[{{Schwarz} {et~al.}(2016){Schwarz}, {Bergin}, {Cleeves}, {Blake},
  {Zhang}, {{\"O}berg}, {van Dishoeck}, \& {Qi}}]{Schwarz2016}
{Schwarz}, K.~R., {Bergin}, E.~A., {Cleeves}, L.~I., {et~al.} 2016, \apj, 823,
  91

\bibitem[{Sheehan(2018)}]{Sheehan2018b}
Sheehan, P. 2018, {psheehan/pdspy: pdspy: A MCMC Tool for Continuum and
  Spectral Line Radiative Transfer Modeling}, , , doi:10.5281/zenodo.2455079

\bibitem[{{Sheehan} \& {Eisner}(2017)}]{Sheehan2017b}
{Sheehan}, P.~D., \& {Eisner}, J.~A. 2017, \apj, 851, 45

\bibitem[{{Simon} {et~al.}(2000){Simon}, {Dutrey}, \& {Guilloteau}}]{Simon2000}
{Simon}, M., {Dutrey}, A., \& {Guilloteau}, S. 2000, \apj, 545, 1034

\bibitem[{{Simon} {et~al.}(2017){Simon}, {Guilloteau}, {Di Folco}, {Dutrey},
  {Grosso}, {Pi{\'e}tu}, {Chapillon}, {Prato}, {Schaefer}, {Rice}, \&
  {Boehler}}]{Simon2017}
{Simon}, M., {Guilloteau}, S., {Di Folco}, E., {et~al.} 2017, \apj, 844, 158

\bibitem[{{Stassun} {et~al.}(2014){Stassun}, {Feiden}, \&
  {Torres}}]{Stassun2014}
{Stassun}, K.~G., {Feiden}, G.~A., \& {Torres}, G. 2014, New Astronomy Reviews,
  60, 1

\bibitem[{{Tazzari} {et~al.}(2017){Tazzari}, {Beaujean}, \&
  {Testi}}]{Tazzari2017}
{Tazzari}, M., {Beaujean}, F., \& {Testi}, L. 2017, {galario: Gpu Accelerated
  Library for Analyzing Radio Interferometer Observations}, Astrophysics Source
  Code Library, , , ascl:1710.022

\bibitem[{{Tazzari} {et~al.}(2018){Tazzari}, {Beaujean}, \&
  {Testi}}]{Tazzari2018}
---. 2018, \mnras, 476, 4527

\bibitem[{{Torres} {et~al.}(2006){Torres}, {Quast}, {da Silva}, {de La Reza},
  {Melo}, \& {Sterzik}}]{Torres2006}
{Torres}, C.~A.~O., {Quast}, G.~R., {da Silva}, L., {et~al.} 2006, \aap, 460,
  695

\bibitem[{{Vrba} {et~al.}(1989){Vrba}, {Rydgren}, {Chugainov}, {Shakovskaia},
  \& {Weaver}}]{Vrba1989}
{Vrba}, F.~J., {Rydgren}, A.~E., {Chugainov}, P.~F., {Shakovskaia}, N.~I., \&
  {Weaver}, W.~B. 1989, \aj, 97, 483

\bibitem[{{Ward-Duong} {et~al.}(2018){Ward-Duong}, {Patience}, {Bulger}, {van
  der Plas}, {M{\'e}nard}, {Pinte}, {Jackson}, {Bryden}, {Turner}, {Harvey},
  {Hales}, \& {De Rosa}}]{WardDuong2018}
{Ward-Duong}, K., {Patience}, J., {Bulger}, J., {et~al.} 2018, \aj, 155, 54

\bibitem[{{Weingartner} \& {Draine}(2001)}]{Weingartner2001}
{Weingartner}, J.~C., \& {Draine}, B.~T. 2001, \apj, 548, 296

\bibitem[{{White} \& {Ghez}(2001)}]{White2001}
{White}, R.~J., \& {Ghez}, A.~M. 2001, \apj, 556, 265

\bibitem[{{Williams} \& {Best}(2014)}]{Williams2014}
{Williams}, J.~P., \& {Best}, W.~M.~J. 2014, \apj, 788, 59

\bibitem[{{Wu} {et~al.}(2017{\natexlab{a}}){Wu}, {Close}, {Eisner}, \&
  {Sheehan}}]{Wu2017c}
{Wu}, Y.-L., {Close}, L.~M., {Eisner}, J.~A., \& {Sheehan}, P.~D.
  2017{\natexlab{a}}, \aj, 154, 234

\bibitem[{{Wu} \& {Sheehan}(2017)}]{Wu2017b}
{Wu}, Y.-L., \& {Sheehan}, P.~D. 2017, \apjl, 846, L26

\bibitem[{{Wu} {et~al.}(2017{\natexlab{b}}){Wu}, {Sheehan}, {Males}, {Close},
  {Morzinski}, {Teske}, {Haug-Baltzell}, {Merchant}, \& {Lyons}}]{Wu2017a}
{Wu}, Y.-L., {Sheehan}, P.~D., {Males}, J.~R., {et~al.} 2017{\natexlab{b}},
  \apj, 836, 223

\bibitem[{{Zacharias} {et~al.}(2003){Zacharias}, {Urban}, {Zacharias},
  {Wycoff}, {Hall}, {Germain}, {Holdenried}, \& {Winter}}]{Zacharias2003}
{Zacharias}, N., {Urban}, S.~E., {Zacharias}, M.~I., {et~al.} 2003, VizieR
  Online Data Catalog, I/289

\end{thebibliography}

\appendix

\section{The MCMC Fitting Procedure}
\label{section:mcmc_fitting}

\begin{figure*}[b!]
\centering
\figurenum{A1}
\includegraphics[width=6.5in]{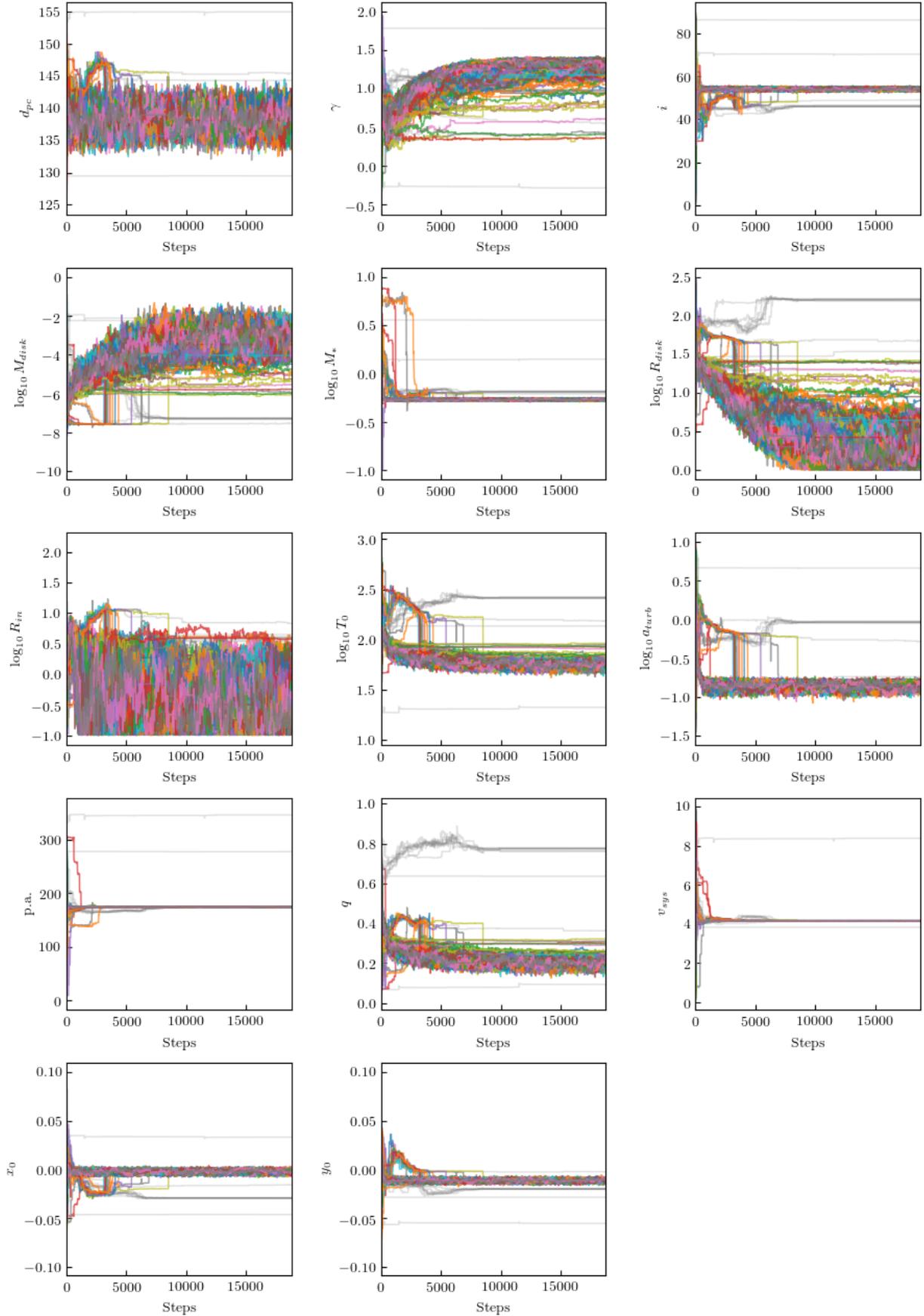}
\caption{Steps taken by the MCMC walkers in the fit to 2M1626--2527. The ``lost" walkers are shown in gray.}
\label{fig:steps_ROXs12}
\end{figure*}

We fit our radiative transfer model, as described in Section \ref{section:modeling}, to our data using the Markov Chain Monte Carlo (MCMC) fitting code \texttt{emcee} \citep{ForemanMackey2013}. \texttt{emcee} uses an implemenation of the \citet{Goodman2010} Affine-Invariant ensemble sampler to sample parameter space with a large number of walkers. We use 200 separate MCMC ``walkers", which we spread out over a large range of parameter space. These walkers are allowed to move through parameter space, with their steps being towards another randomly selected walker and the size of the step dictated by a comparison of the goodness of fit of each set of walker parameters. The likelihood function for a given set of parameters $\hat{\theta}$ used for the fit is given by
\begin{equation}
\ell(\hat{\theta}) = \exp\left[-\frac{1}{2}\sum_{i=1}^{N} \sum_{j=1}^{M} \left|V(u_i, v_i, \nu_j) - M(u_i, v_i, \nu_j \, | \, \hat{\theta}) \right|^2 \, W(u_i, v_i, \nu_j)  \right],
\end{equation}
where $V(u_i, v_i, \nu_j)$ are the complex visibility data at baseline $(u_i, v_i)$ and frequency $\nu_j$, $W(u_i, v_i, \nu_j)$ are the visibility data weights, and $M(u_i, v_i, \nu_j \, | \, \hat{\theta})$ is the model with parameters $\hat{\theta}$ evaluated at baseline $(u_i, v_i)$ and frequency $\nu_j$. $N$ is the total number of baselines in the visibility data, and $M$ is the number of channels included in the fit. We consider a fit to be ``converged" when the median walker position does not change significantly compared to the spread of the walkers over a significant number of steps. We show the steps taken by the walkers in each fit in Figures \ref{fig:steps_ROXs12}, \ref{fig:steps_CTCha}, and \ref{fig:steps_DHTau}. 

We note that while we commonly refer to the 14 parameters of our model fit as $\hat{\theta} = \{M_{*}, M_{disk}, R_{in}, R_{disk},$ $\gamma, T_0, q, \xi, v_{sys}, i, d_{pc}, P.A., x_0, y_0\}$, reasonable ranges of values for a number of these parameters span multiple orders of magnitude. For such parameters, it makes practical sense to use the log of the parameter as the actual fit quantity. For this reason, the actual parameters being fit are $\hat{\theta} = \{\log_{10}{M_{*}}, \log_{10}{M_{disk}}, \log_{10}{R_{in}}, \log_{10}{R_{disk}}, \gamma, \log_{10}{T_0}, q, \log_{10}{\xi},$ $v_{sys}, i, d_{pc}, P.A., x_0, y_0\}$. As the linear values are more intuitive to understand however, for most tables and figures and discussion in the text we convert the posterior Markov Chain Monte Carlo samples to linear values before evaluating best fit values. We assume a uniform prior on all parameters except for the source distance ($d_{pc}$) and stellar mass ($M_*$), with the following limits: $-2 \leq \log_{10}{M_{star}} \leq 1$, $-10 \leq \log_{10}{M_{disk}} \leq 0$, $-1 \leq \log_{10}{R_{in}} \leq 1.5$, $0 \leq \log_{10}{R_{disk}} \leq 2.5$, $-0.5 \leq \gamma \leq 2$, $1 \leq \log_{10}{T_0} \leq 3$, $0 \leq q \leq 1$, $-1.5 \leq \log_{10}{a_{turb}} \leq 1$, $0^{\circ} \leq i \leq 90^{\circ}$, $0^{\circ} \leq p.a. \leq 360^{\circ}$, $-0.1" \leq x_0 \leq 0.1"$, $-0.1" \leq y_0 \leq 0.1"$. In principle, sensitive ALMA spectral line data can distinguish between inclinations above and below $90^{\circ}$ \citep[e.g.,][]{Czekala2015}. However our initial tests found that the data does not have the required sensitivity, with the walkers split approximately evenly between the two solutions, so we limit $i < 90^{\circ}$. 

For a young star for which the stellar mass is not a priori known, the probability of that star having mass $M_*$ is given by the stellar initial mass function. As such we use the Chabrier IMF \citep{Chabrier2003} as a prior on the stellar mass in our fit. Furthermore, while $d_{pc}$ is left as a free parameter in the fit, our ALMA data do not provide a direct constraint on its value. As stellar mass is highly degenerate with source distance \citep[e.g.,][]{Czekala2015}, placing some constraint on source distance is important. Moreover, strong constraints on source distance lead to higher precision measurements of stellar mass. Here, we use direct measurements of source distances from trigonometric parallaxes by $Gaia$ to place a prior constraint on the distance. We find from $Gaia$ that 2M1626--2527 has a parallax of $7.232 \pm 0.074$ mas and a distance of $138.258^{+1.426}_{-1.396}$ pc, CT Cha has a parallax of $5.214 \pm 0.021$ mas and a distance of $191.775^{+0.781}_{-0.774}$ pc, and DH Tau has a parallax of $7.388 \pm 0.069$ mas and a distance of $135.355^{+1.282}_{-1.258}$ pc \citep{GaiaCollaboration2018}. We assume a Gaussian prior on the parallax, with a mean given by the measured $Gaia$ parallax and the standard deviation given by the $Gaia$ uncertainty. The model parallax is calculated as $1/d_{pc}$.

While in general the walkers tend to converge into one bunch, a handful of walkers appear to become lost in parameter space and have trouble finding their way to the main group (e.g., see Figure \ref{fig:steps_ROXs12}). We believe that this is the result of how the walkers move through parameter space. To determine the next step for a walker, a different walker is randomly drawn from the sample of walkers. The walker will (randomly) move towards or away from the randomly selected walker with a step size between 1/a and a times the distance between them (a = 2 by default in \texttt{emcee}). The walkers become ``lost", however, because the global minimum is quite narrow in some parameters (e.g., $v_{sys}$), so the likelihood that the step puts it into the main bunch (i.e. a = 1) gets very small. It therefore takes a large number of step proposals before the walker finally makes the jump. While we have confidence that, if given enough time, these walkers would converge as well, the computational demands of these fits are significant, so we opt not to continue running the fits. Instead, we discard any walkers with $(\chi^2 - \chi^2_{min}) > 44$, appropriate for a $\chi^2$ distribution with 14 parameters and $p = 0.00006$ ($\sim4\sigma$). The autocorrelation time for these fits is typically about $\sim1000$ steps, so with $5000$ post-burn-in steps in the chain, we have $\sim5$ independent sets of walker positions. With 200 walkers,
%and typically $\sim500$ steps post-burn-in, 
or a total of $\sim1000$ independent samples, we would expect $\ll1$ ($\sim0.06$) walkers to have $(\chi^2 - \chi^2_{min}) > 44$, so this cut will identify any ``lost" walkers that are not likely representative of the underlying distribution.
%As we find that the ``lost" walkers do not significantly affect our results, we opt to not continue running the fits until all walkers are converged. 
We show the steps taken by all walkers in Figures \ref{fig:steps_ROXs12}, \ref{fig:steps_CTCha}, and \ref{fig:steps_DHTau}. The discarded walkers shown in gray to demonstrate that the cuts are reasonable. %so we use sigma-clipping on the post-burn-in walker positions to trim the walkers that exhibit this lost behavior. The trimmed walkers are shown in gray in Figures \ref{fig:steps_ROXs12}, \ref{fig:steps_CTCha}, and \ref{fig:steps_DHTau} to demonstrate that we are making reasonable cuts.

\begin{figure*}[t]
\centering
\figurenum{A4}
\includegraphics[width=7in]{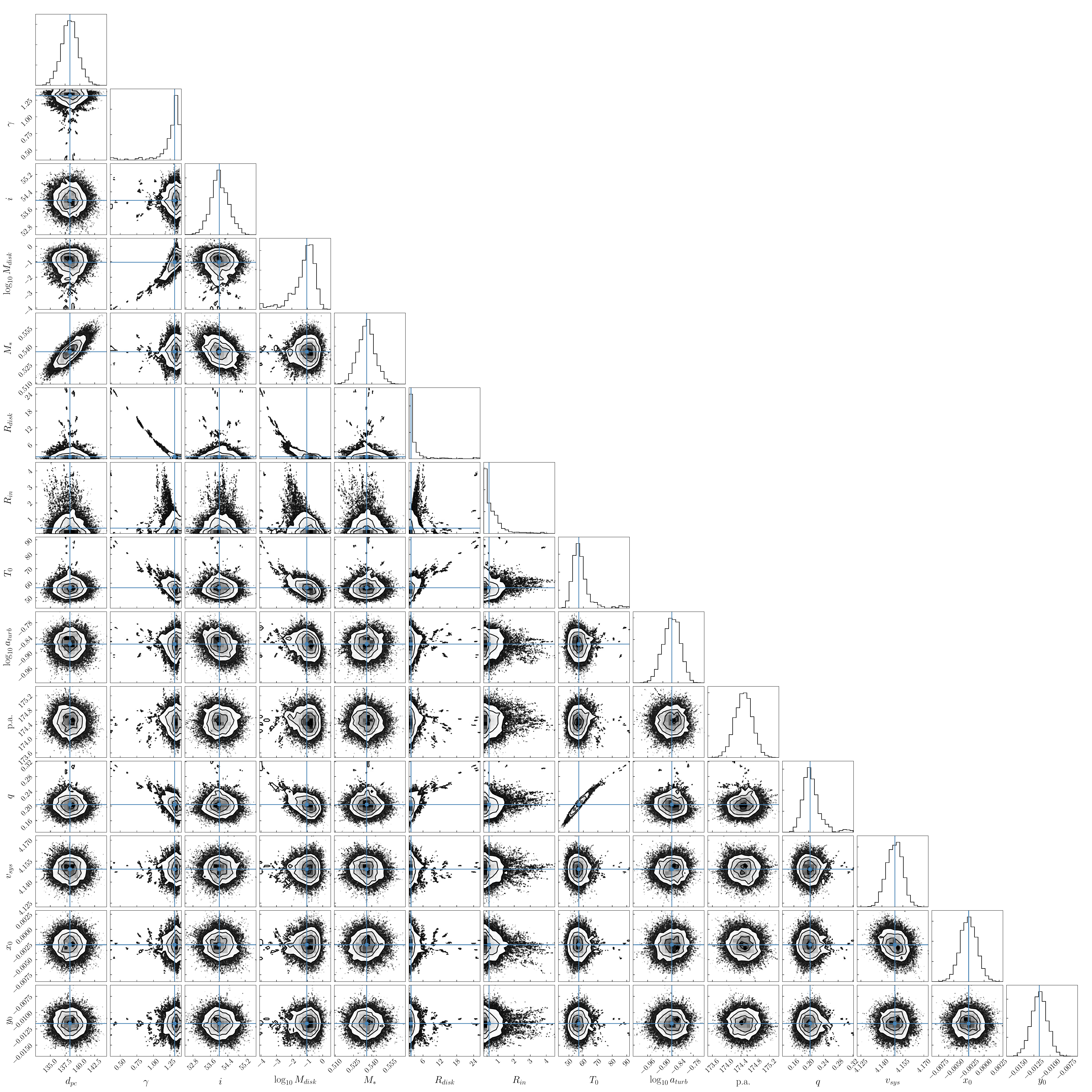}
\caption{One- and two-dimensional posterior probability density functions for the fit to 2M1626--2527. The best fit values are shown as horizontal and vertical lines.}
\label{fig:triangle_ROXs12}
\end{figure*}

\begin{figure*}[t]
\centering
\figurenum{A5}
\includegraphics[width=7in]{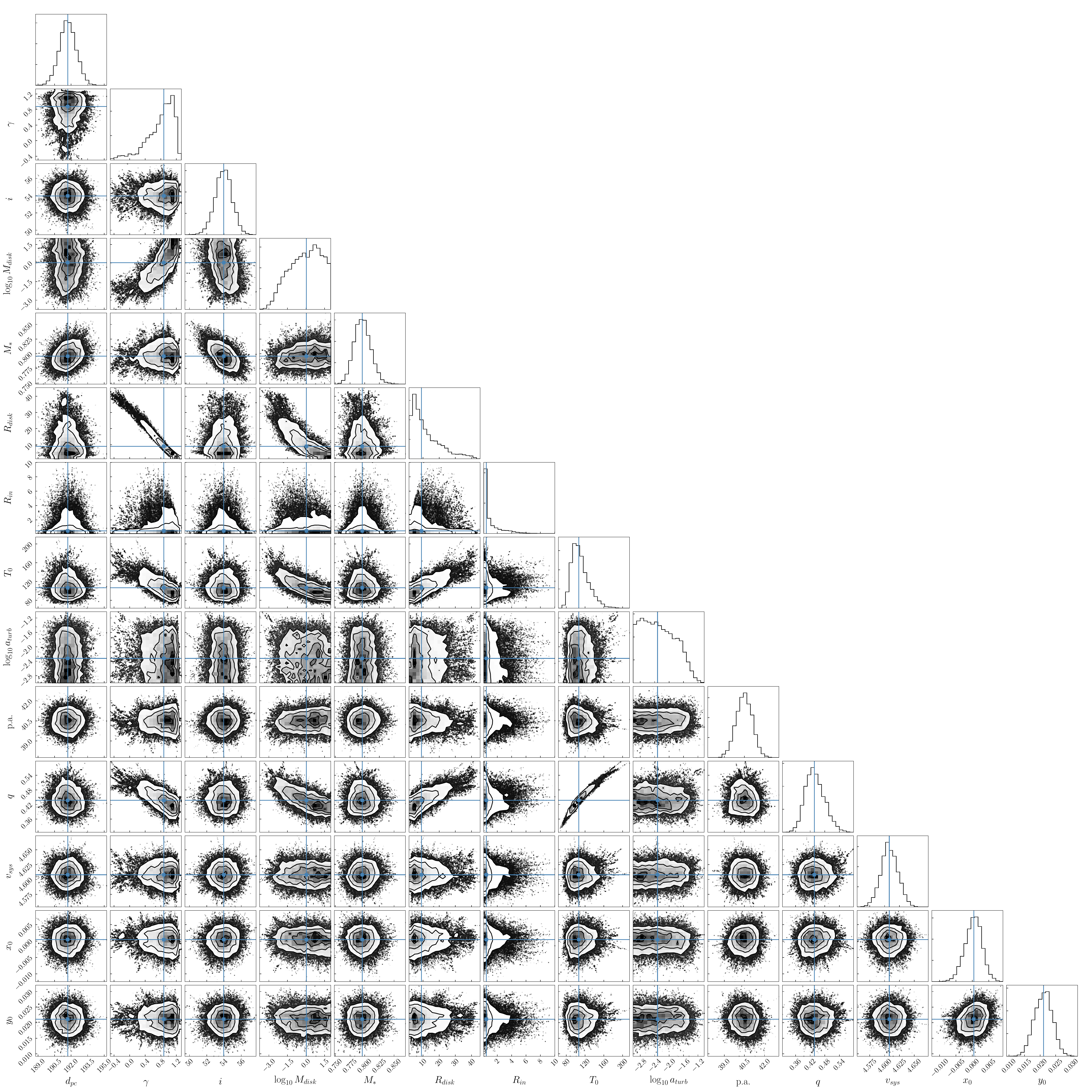}
\caption{One- and two-dimensional posterior probability density functions for the fit to CT Cha. The best fit values are shown as horizontal and vertical lines.}
\label{fig:triangle_CTCha}
\end{figure*}

\begin{figure*}[t]
\centering
\figurenum{A6}
\includegraphics[width=7in]{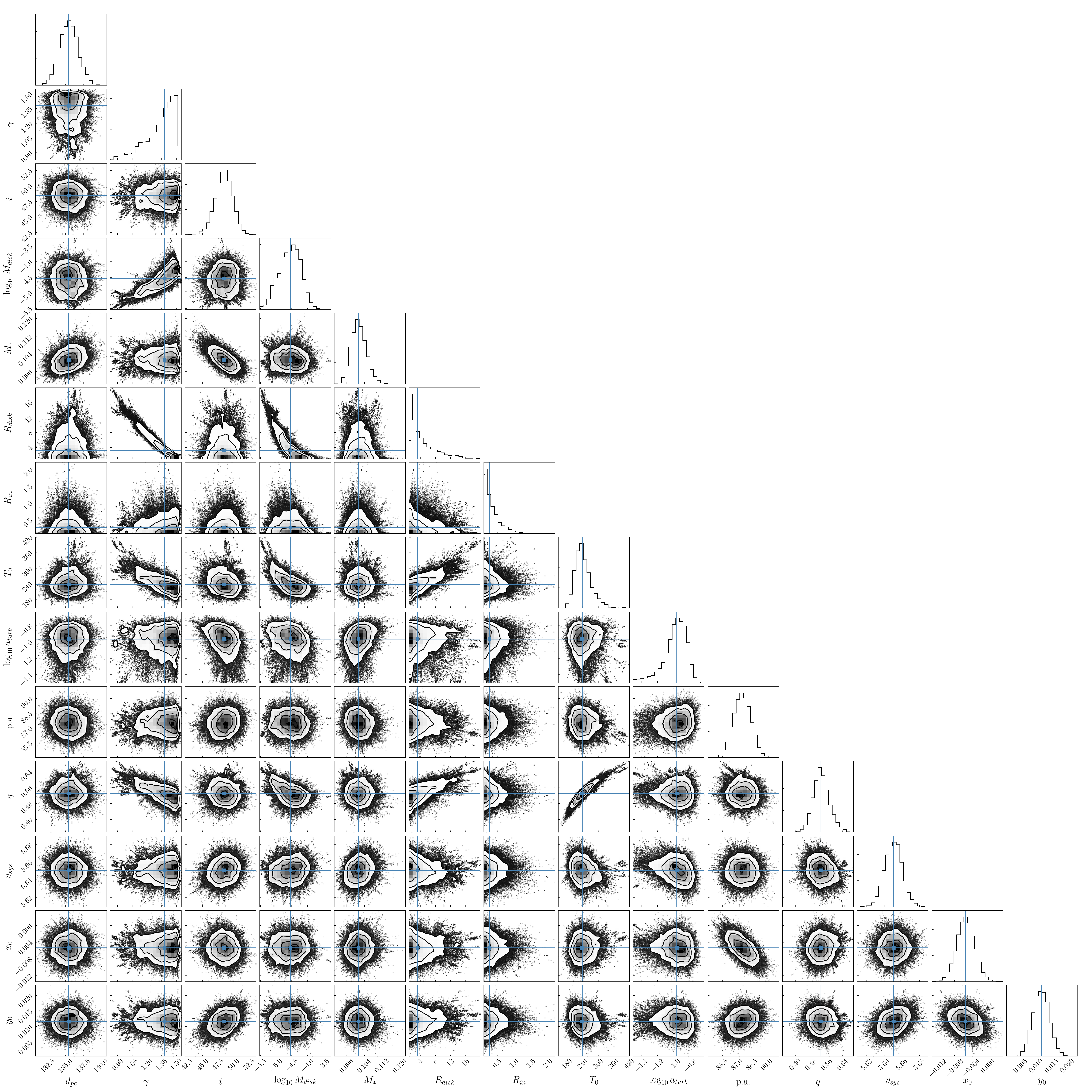}
\caption{One- and two-dimensional posterior probability density functions for the fit to DH Tau. The best fit values are shown as horizontal and vertical lines.}
\label{fig:triangle_DHTau}
\end{figure*}

In Figures \ref{fig:triangle_ROXs12}, \ref{fig:triangle_CTCha}, and \ref{fig:triangle_DHTau} we show the two dimensional posterior probability density functions for each pair of parameters, after the burn-in steps and ``lost" walkers have been discarded, for each source. %As the ``lost" walkers make it difficult to see the distribution of the main bunch, we use sigma clipping to discard them and zoom in on the main group. 
%The clipped walkers are shown in gray in Figures \ref{fig:steps_ROXs12}, \ref{fig:steps_CTCha}, and \ref{fig:steps_DHTau}. 
We estimate the best fit value for a parameter as the median of the walker positions in that parameter after convergence is achieved, the ``lost" walkers are discarded, and the walker steps are trimmed by the autocorrelation time. Similarly, we estimate the uncertainties on that value as the region around the median containing 68\% of the walker positions after convergence and ``lost" walkers are discarded. For parameters where the walkers move up against an edge of the allowed range ($a_{turb}$ for CT Cha and DH Tau, $R_{disk}$ for DH Tau and 2M1626--2527, $M_{disk}$ for CT Cha, $R_{in}$ for all), we instead use the value containing 99.7\% of walkers above or below that value as a limit.

%\section{AK Sco as a Test Case}

%To test the \texttt{pdspy} code and ensure it is working properly, we use it to model the AK Sco data from \citet{Czekala2015}, and compare with the results from that work using the \texttt{DiskJockey} code. AK Sco is a double lined pre-main sequence eclipsing binary that has a precise measurement of stellar mass (up to an unknown inclination) from radial velocity variations \citep{Alencar2003}. \citet{Czekala2015} used CO 2--1 ALMA observations to directly measure the stellar mass independently, and found that if the stellar orbital inclination is the same as the disk inclination, the two measurements differ by $<1\%$. The excellent agreement between the two measurements provided a validation of using disk kinematics to measure stellar masses. With \texttt{pdspy} we find best fit parameters for AK Sco of: $d_{pc} = 140.7^{+  1.7}_{-  0.4}$, $i$ = $109.5^{+ 0.3}_{- 0.3}$, $M_{disk}$ = $-3.42^{+0.09}_{-0.09}$, $M_{*}$ = $2.446^{+0.028}_{-0.015}$, $R_{disk}$ = $ 15.5^{+  0.5}_{-  0.5}$, $T_0$ = $  304^{+   10}_{-   11}$, $a_{turb}$ = $ 0.297^{+ 0.014}_{- 0.013}$, p.a. = $ 50.9^{+  0.2}_{-  0.2}$, $q$ = $0.51^{+0.01}_{-0.01}$, $v_{sys}$ = $-26.082^{+0.009}_{-0.009}$, $x_0$ = $-0.039^{+0.001}_{-0.002}$, and $y_0$ = $-0.040^{+0.001}_{-0.001}$. These values are all in excellent agreement with what is found by \citet{Czekala2015}, confirming that the \texttt{pdspy} code is working correctly.

\section{Additional Model Parameters}
\label{section:model_parameters}

\texttt{pdspy} is designed to be a flexible modeling tool that can handle a range of input parameters, priors, and models. For example, here we have left the surface density profile power law exponent, $\gamma=1$, as a free parameter. However, the walker plots show that $\gamma$ converges slowly and does not ultimately affect the measured value of $M_*$. As such, it may be wise to fix $\gamma=1$ \citep[e.g.,][]{Czekala2015} to reduce the needed computational time, and this can easily be done in $pdspy$ by changing a single boolean value in the configuration file. This, of course, may not be wise if the goal is to measure disk structure, as parameters like $M_{disk}$ and $R_{disk}$ are degenerate with $\gamma$, but for studies measuring stellar mass this would be appropriate.
%In practice, however, $\gamma$ can easily be made to vary in \texttt{pdspy}.}
In addition to adding or removing parameters from our model, \texttt{pdspy} also includes a variety of models that can be employed. Here we have employed an exponentially tapered surface density profile, however \texttt{pdspy} includes options for a truncated disk model as well. Moreover, up to three gaps and a cavity can be added as features of the model. Finally, simple Gaussian priors can be added to any parameter, and a few ``special" priors are included in the code. As described above the distance can be constrained with a prior on the parallax, and the stellar mass can be constrained by using the IMF as a prior. 

In all, \texttt{pdspy} currently has over 25 parameters that can be turned on or off at the users discretion, as well as options to use a truncated disk and to include continuum opacity/subtraction. We are also exploring adding additional functionality, including non-vertically isothermal disk temperatures and inclusion of envelope emission for embedded protostars. More information, including the current, up-to-date status of modeling options, can be found at https://github.com/psheehan/pdspy.

\section{Channel Map Images and Models}
\label{section:channel_maps}

\begin{figure*}[t]
\centering
\figurenum{C1}
\includegraphics[width=7in]{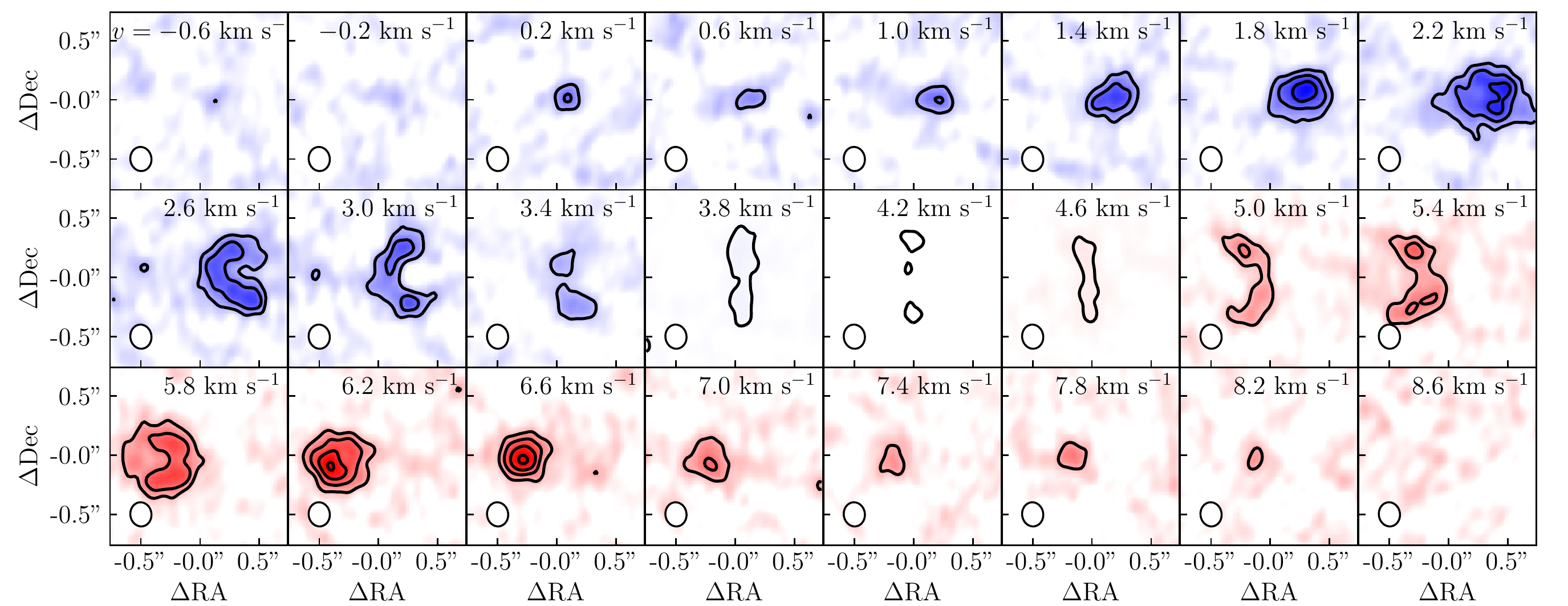}
\includegraphics[width=7in]{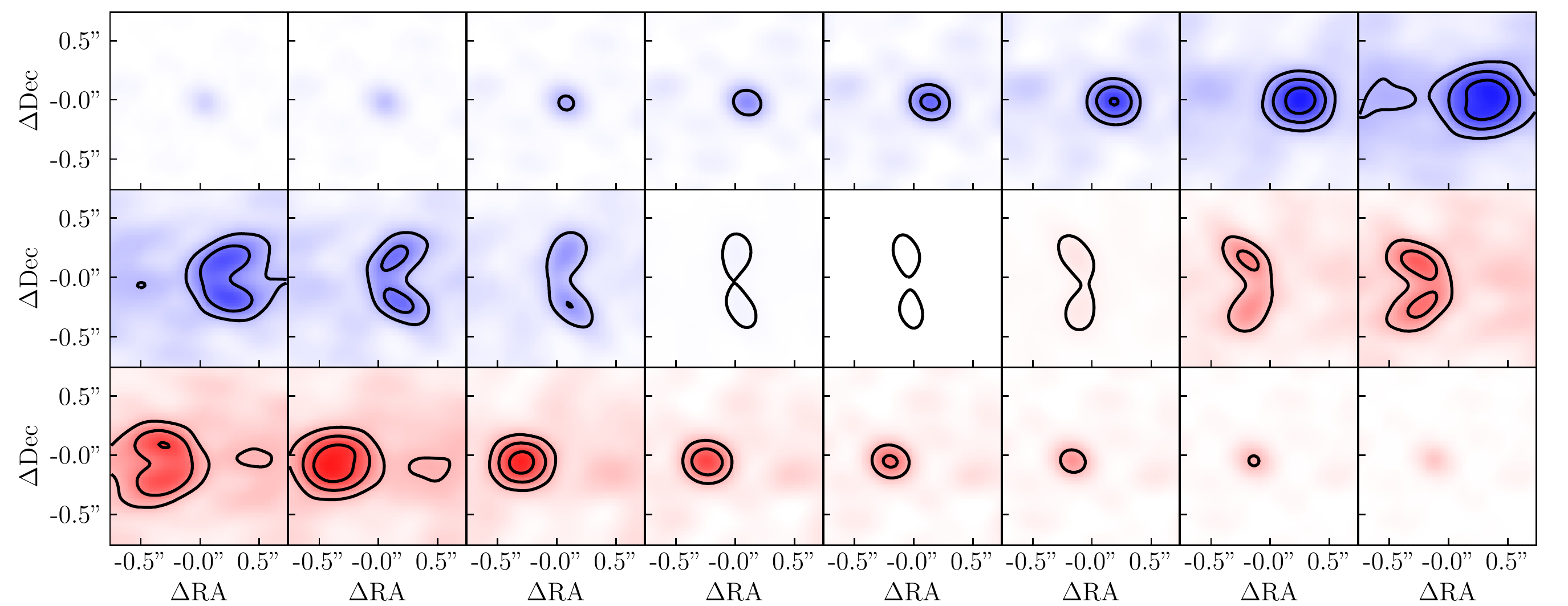}
\includegraphics[width=7in]{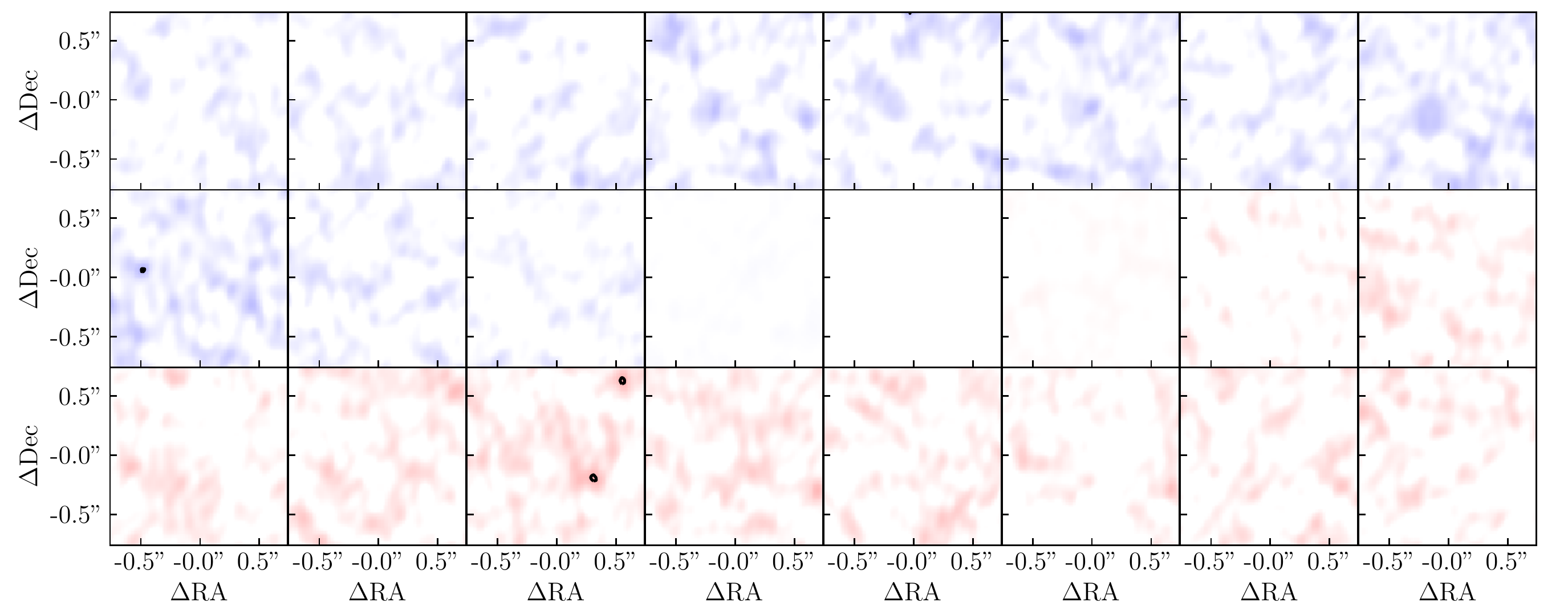}
\caption{Modeling results for 2M1626--2527. The top plot shows our ALMA CO (2--1) channel maps, with contours starting at 3.5$\sigma$ and continuing in increments of 3$\sigma$. The second plot shows the best fit model channel map images, and the third plot shows the residuals, calculated in the Fourier plane and then Fourier transformed to produce an image.}
\label{fig:model_ROXs12}
\end{figure*}

\begin{figure*}[t]
\centering
\figurenum{C2}
\includegraphics[width=7in]{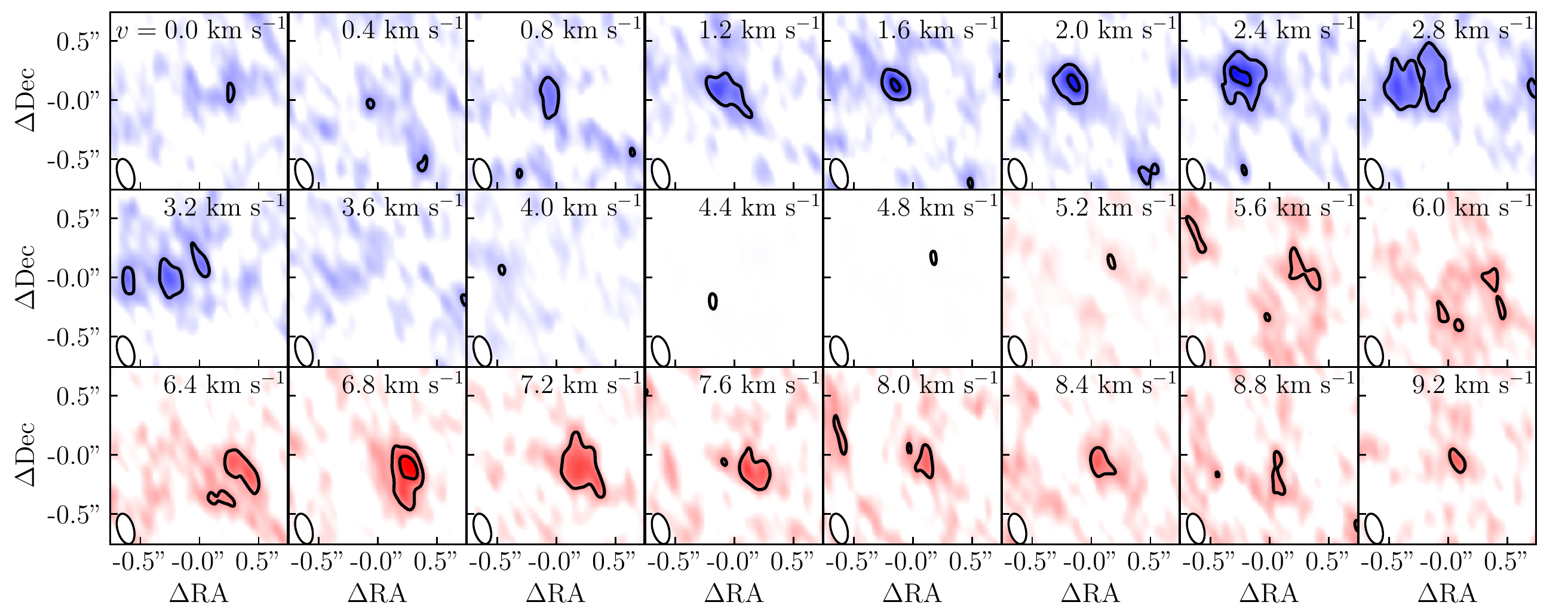}
\includegraphics[width=7in]{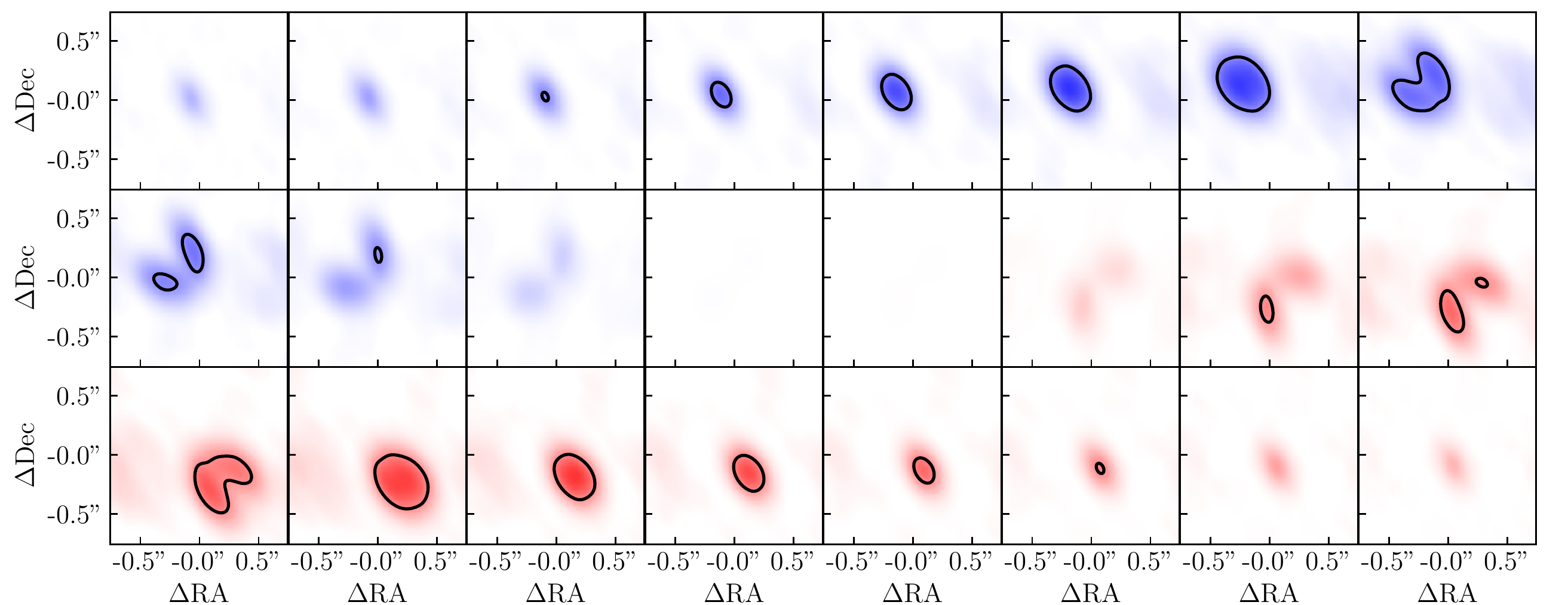}
\includegraphics[width=7in]{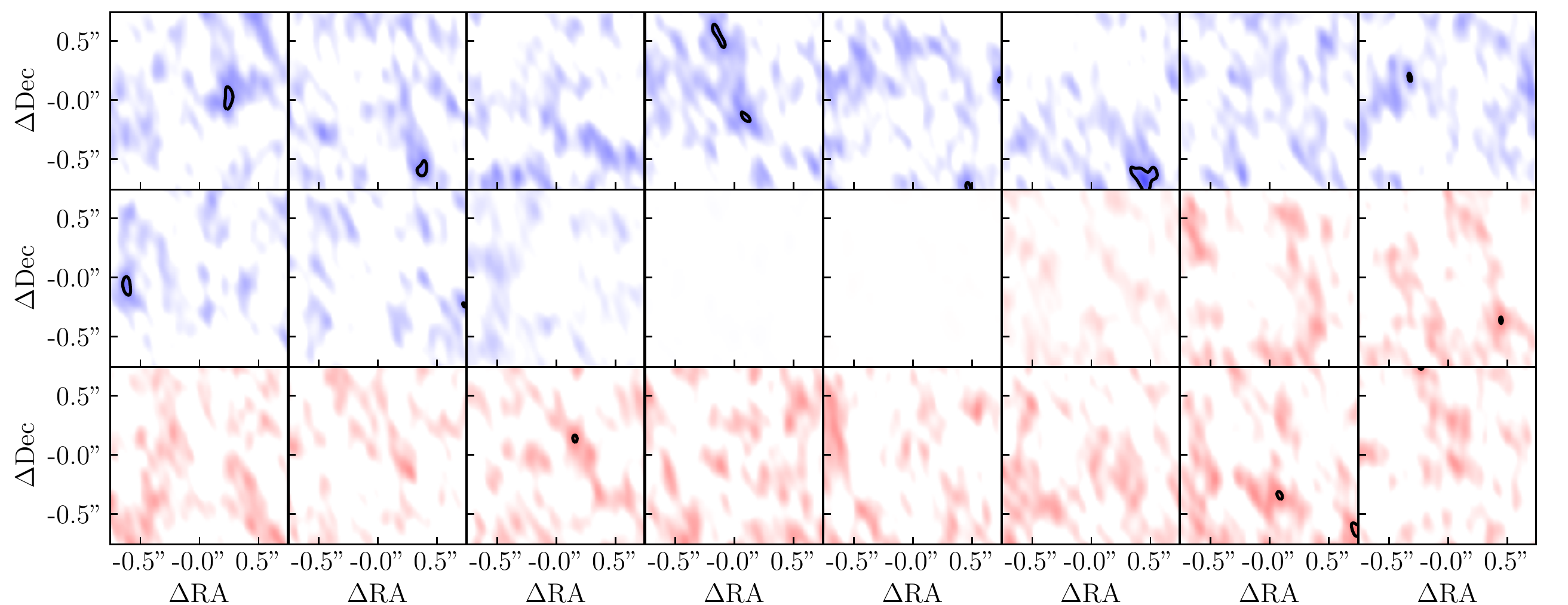}
\caption{Modeling results for CT Cha. The top plot shows our ALMA CO (2--1) channel maps, with contours starting at 3.5$\sigma$ and continuing in increments of 3$\sigma$. The second plot shows the best fit model channel map images, and the third plot shows the residuals, calculated in the Fourier plane and then Fourier transformed to produce an image.}
\label{fig:model_CTCha}
\end{figure*}

\begin{figure*}[t]
\centering
\figurenum{C3}
\includegraphics[width=5.5in]{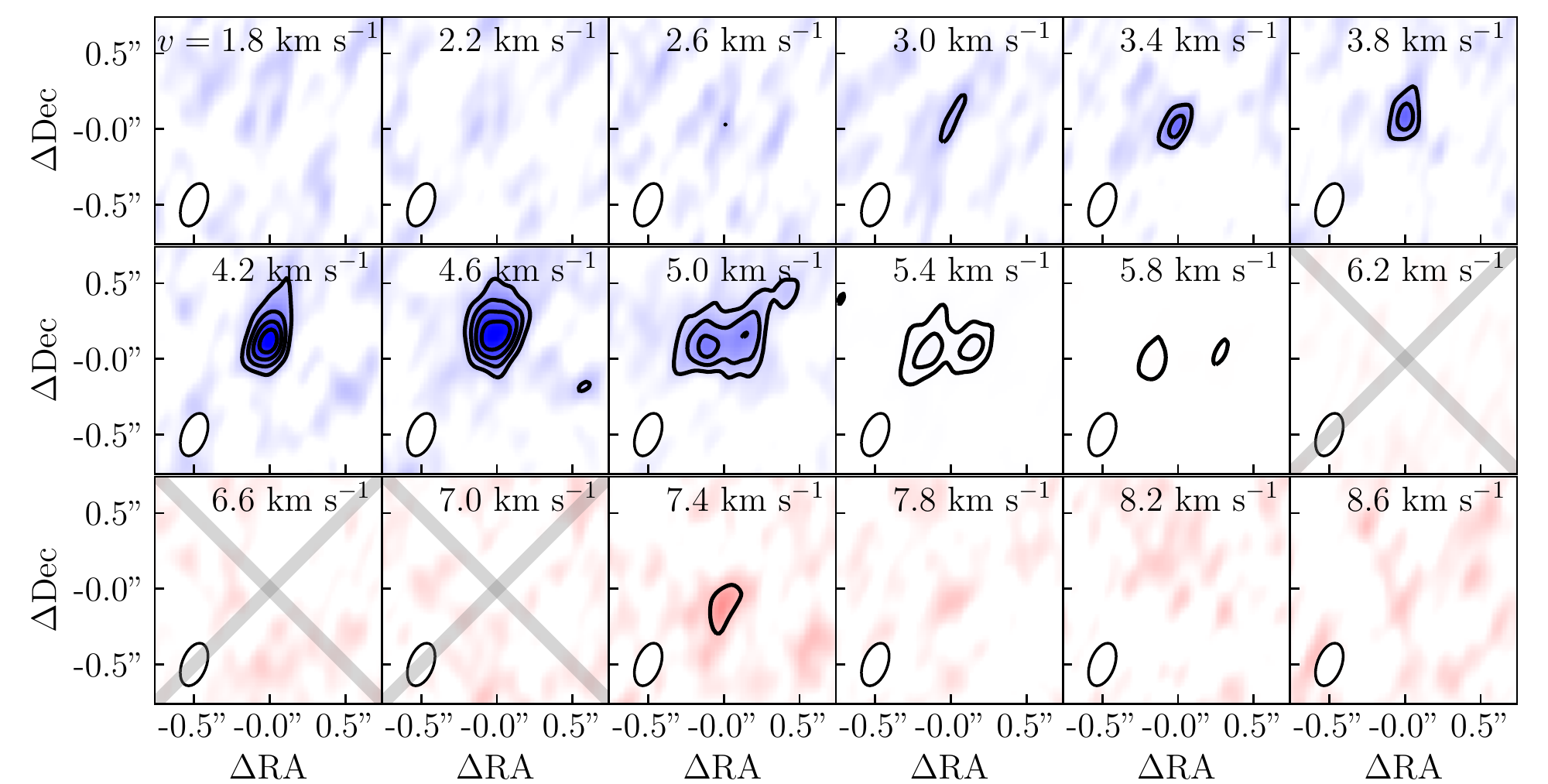}
\includegraphics[width=5.5in]{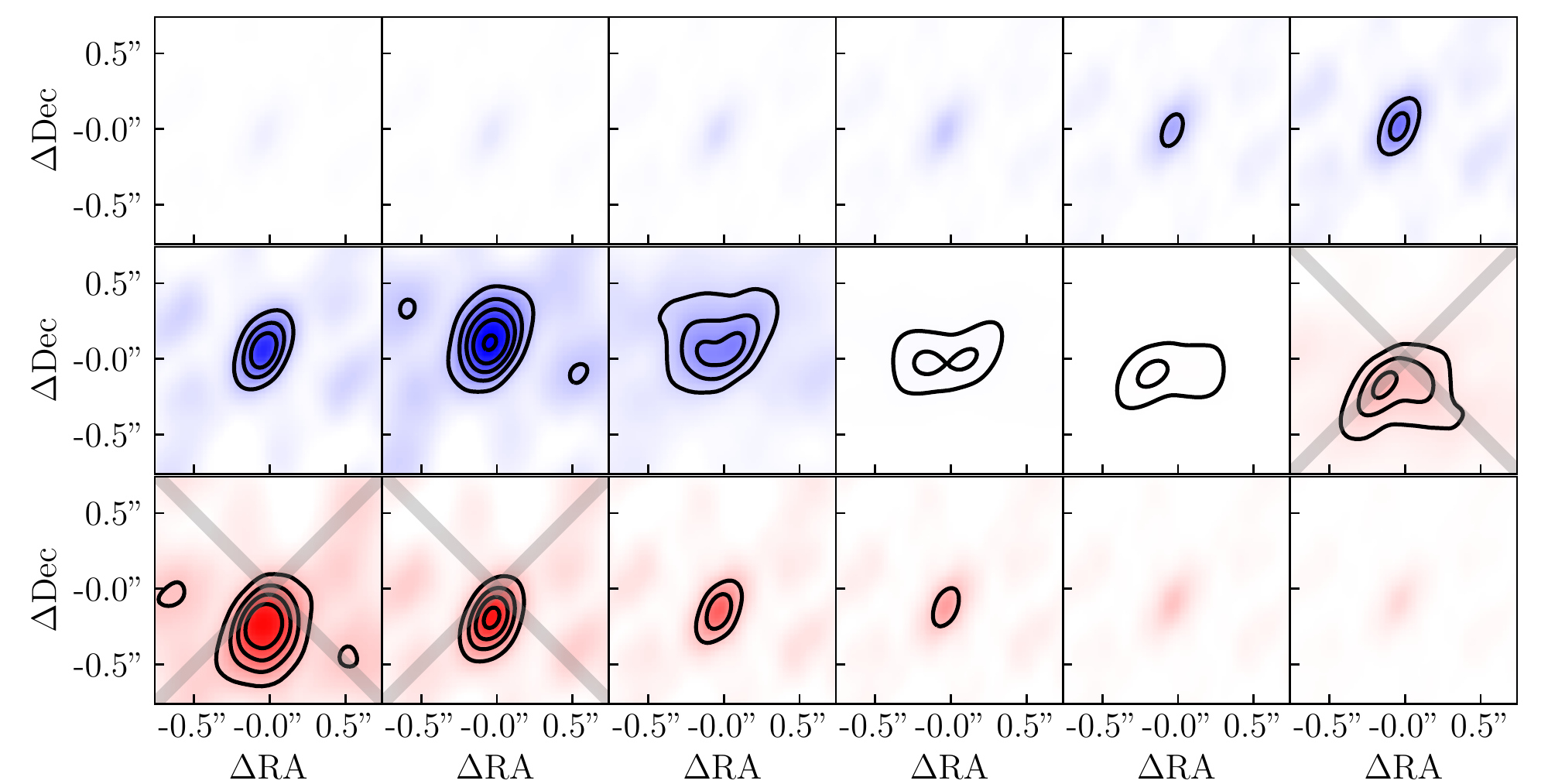}
\includegraphics[width=5.5in]{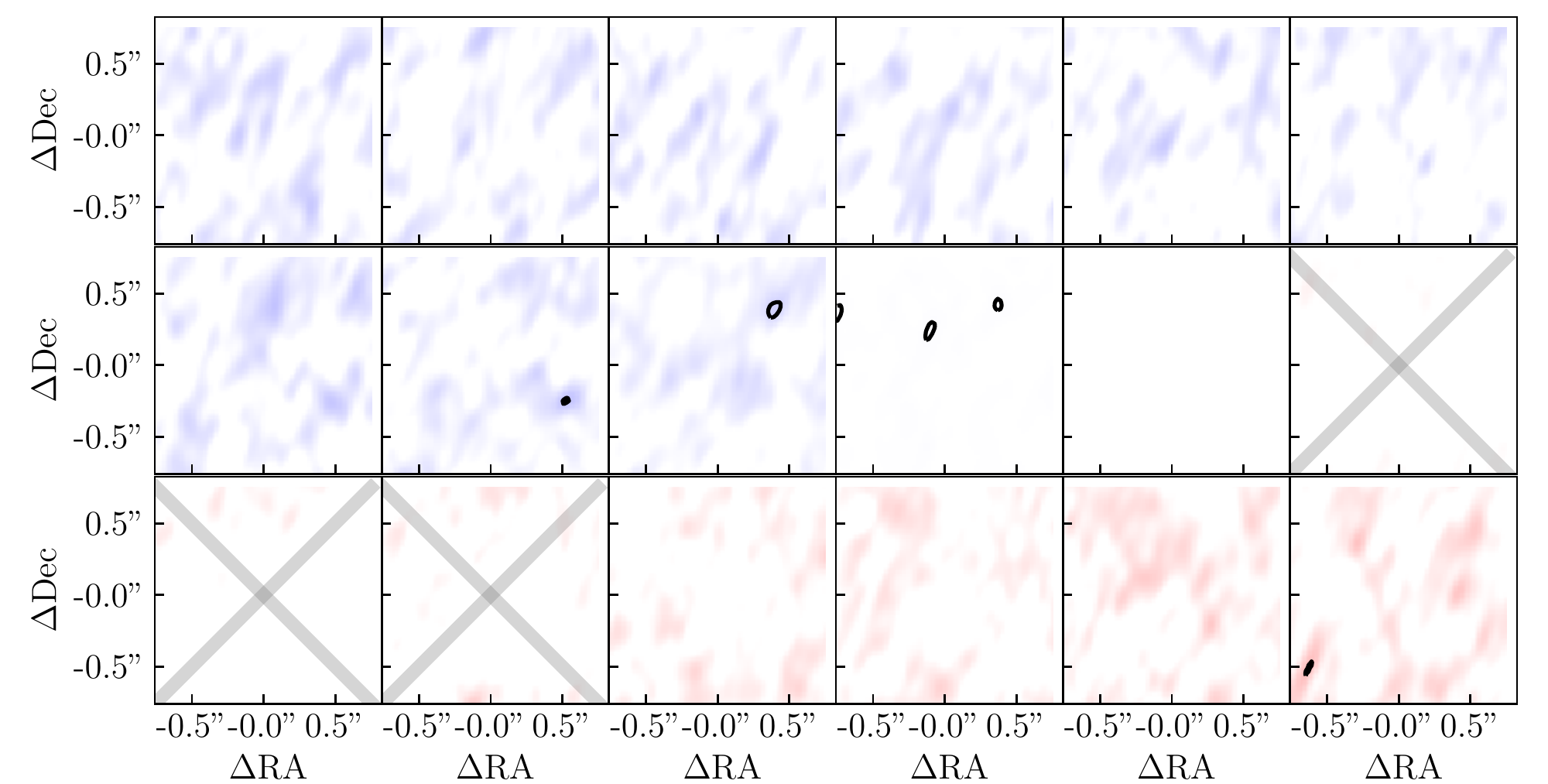}
\caption{Modeling results for DH Tau. The top plot shows our ALMA CO (2--1) channel maps, with contours starting at 3.5$\sigma$ and continuing in increments of 3$\sigma$. The second plot shows the best fit model channel map images, and the third plot shows the residuals, calculated in the Fourier plane and then Fourier transformed to produce an image. The channels with marked with an ``X" show a lack of redshifted emission, likely due to absorption by a foreground cloud. The ``X" indicates that these channels were excluded from our modeling.}
\label{fig:model_DHTau}
\end{figure*}

In Figures \ref{fig:model_ROXs12}, \ref{fig:model_CTCha}, and \ref{fig:model_DHTau} we show the best fit models and residuals for each of our sources, here including every channel in the maps.

\begin{figure*}[t]
\centering
\figurenum{A2}
\includegraphics[width=6.5in]{figureA2.pdf}
\caption{Steps taken by the MCMC walkers in the fit to CT Cha. The ``lost" walkers are shown in gray.}
\label{fig:steps_CTCha}
\end{figure*}

\begin{figure*}[t]
\centering
\figurenum{A3}
\includegraphics[width=6.5in]{figureA3.pdf}
\caption{Steps taken by the MCMC walkers in the fit to DH Tau. The ``lost" walkers are shown in gray.}
\label{fig:steps_DHTau}
\end{figure*}

\end{document}